\documentclass{article}

\usepackage{arxiv}

\usepackage[utf8]{inputenc} 
\usepackage[T1]{fontenc}    
\usepackage{hyperref}       
\usepackage{url}            
\usepackage{booktabs}       
\usepackage{amsfonts}       
\usepackage{nicefrac}       
\usepackage{microtype}      
\usepackage{lipsum}
\usepackage{graphicx}
\usepackage[justification=centering]{caption}
\usepackage{subfigure} 
\usepackage{enumerate}
\usepackage{amsmath}
\usepackage{booktabs}
\usepackage{longtable}
\usepackage{float} 

\title{A novel design of industrial real-time CT system based on sparse-view reconstruction and deep-learning image enhancement}

\author{
 Zheng Fang \\
  School of Aerospace Engineering\\
  Xiamen University\\
  Xiamen, 361102, Fujian, China \\
  \texttt{fangzheng@xmu.edu.cn} \\
   \And
 Tingjun Wang \\
  School of Aerospace Engineering\\
  Xiamen University\\
  Xiamen, 361102, Fujian, China \\
  \texttt{vintagewtj@163.com} \\
  \And
 Bingan Yuan \\
  School of Aerospace Engineering\\
  Xiamen University\\
  Xiamen, 361102, Fujian, China \\
  \texttt{binganyuan@163.com} \\
  \And
 Xinlin Qing \\
  School of Aerospace Engineering\\
  Xiamen University\\
  Xiamen, 361102, Fujian, China \\
  \texttt{XMUqingxinlin@163.com} \\
  \And
 Shunren Li \\
  ASR Technology, Co., Ltd.\\
  Xiamen, 361022, Fujian, China \\
  \texttt{shunrenlixray@163.com} \\  
}

\begin{document}
\maketitle
\begin{abstract}
Industrial CT is useful for defect detection, dimensional inspection and geometric analysis. While it does not meet the needs of industrial mass production, because of its time-consuming imaging procedure. This article proposes a novel stationary real-time CT system with multiple X-ray sources and detectors, which is able to refresh the CT reconstructed slices to the detector frame frequency. This kind of structure avoids the movement of the X-ray sources and detectors. Projections from different angles can be acquired with the objects’ translation, which makes it easier to be integrated into pipeline. All the detectors are arranged along the conveyor, and observe the objects in different angle of view. With the translation of objects, their X-ray projections are obtained for CT reconstruction. To decrease the mechanical size and reduce the number of X-ray sources and detectors, the FBP reconstruction algorithm was combined with deep-learning image enhancement. Medical CT images were applied to train the deep-learning network for its quantity advantage in comparison with industrial ones. It is the first time to adopt this source-detector layout strategy. Data augmentation and regularization were used to elevate the generalization of the network. Time consumption of the CT imaging process was also calculated to prove its high efficiency. It is an innovative design for the 4th industrial revolution, providing an intelligent quality inspection solution for digital production. 
\end{abstract}


\section{Introduction}
X-ray technology has developed quickly since Roentgen discovered the X-ray in 1895 \cite{bib1}. It has many successful applications in a wide range of fields and continue to advance nowadays. X-ray radiology is a universal non-destructive testing approach. In traditional fluoroscopy process, internal structures of a three-dimensional object are compressed into a two-dimensional image along the direction of X-rays, resulting in overlapping of all structures within the object, the clarity in the region of interest is greatly reduced. Although traditional perspective imaging technology has achieved some successes in generating images of clear plane of interest, it does not increase the contrast between different substances in the object, nor dose it fundamentally remove other structures other than the focal plane, thus significantly impairing image quality. Modern tomography, known as CT, was invented by Hounsfield \cite{bib2}, it is a method of reconstructing tomography images based on projection data (sinograms) from multiple angles around the inspected object. During the imaging process, each reconstruction slice is independent from each other, so the interference of different slices in traditional perspective imaging process is fundamentally eliminated, and the structural contrast is significantly enhanced; at the same time, gray value of the image pixels in the tomographic image can truly correspond with the material density of the measured object, and a small density change inside the object can be easily detected and located, which is crucial for industrial non-destructive testing. In fact, the low-contrast detect ability is a key difference between CT and X-ray technologies. This is also the most important reason why CT is rapidly popularized in the field of industrial testing. Despite that the imaging quality of CT is outstanding, the inspection speed of traditional CT is generally slow. Several works have been done to elevate the imaging speed:\cite{bib3}\cite{bib4}\cite{bib5}\cite{bib6}\cite{bib7}.

In general, there are two main solutions to increase the imaging speed, either through hardware improvements or through advanced reconstruction algorithms. When it comes to hardware, the parallel computing ability is crucial. Such computing unit with parallel computing ability like FPGA and GPU can significantly increase the imaging speed. When it comes to algorithms, deep-learning neural network is now applying to analytical or iteration algorithms to increase the computing speed while maintaining the reconstruction quality \cite{bib8}\cite{bib9}\cite{bib10}\cite{bib11}.

The development of CT imaging mode has led to the development of CT image reconstruction algorithms. Traditional CT reconstruction algorithms include two main categories: analytical reconstruction algorithms and iterative reconstruction algorithms. Compared with the iterative algorithm, analytical algorithm has the characteristics of fast reconstruction speed, simple error analysis and small occupation of computing resources, it is the mainstream algorithm applied in the current CT system. In recent years, with the development of deep-learning, its application in CT reconstruction draws people’s attention. Much fewer projections are needed for reconstruction through a well-trained neural network, it will greatly shorten the reconstruction time and provide a solution for real-time CT.

In this paper, we proposed a novel structure of industrial CT inspection system, it has the advantage of acquiring high quality projections in multiple angels for CT reconstruction and highly integrated into the production line, so that the inspection procedure can be done on a conveyor belt. The whole production efficiency will improve greatly by this structure.

Many works have been done to apply the X-ray technology on industrial inspection. X-ray tomography has been a typical technology in non-destructive testing area. Detailed studies of classical tomography development can be found elsewhere \cite{bib12}\cite{bib13}\cite{bib14}\cite{bib15}\cite{bib16}\cite{bib17}. Over the past half century, with further developments in electronics and computers, X-ray computed tomography (CT) achieved rapid development. in early 1970s, CT was used merely in medical field, but it started to appear in industrial field with the evolution of industrial non-destructive technology. Gilboy(et.al) presented the applications of the X-ray technique in industrial area \cite{bib18}. Reimers P(et.al) discussed the development of non-destructive testing (NDT) with X-ray computed tomography \cite{bib19}. Kress J W(et.al) designed an X-ray radiographic system with the ability of three-dimensional tomographic reconstruction \cite{bib20}. Oster R(et.al) demonstrated the usage of CT for optimizing the manufacturing techniques with some supportive experiments \cite{bib21}. The development of deep learning technology has brought a revolutionary impact on the field of industrial automation, many application scenarios have emerged: \cite{bib22}\cite{bib23}\cite{bib24}\cite{bib25}\cite{bib26}.

With the in-depth application of CT technology in the fields of three-dimensional imaging, medical navigation, and rapid security inspection, new requirements such as low-dose imaging, quantitative imaging, and rapid imaging have also put forward higher requirements for CT imaging technology, but the development of standard scanning trajectory CT systems represented by circular trajectory scanning and spiral trajectory scanning has encountered bottlenecks in terms of faster imaging speed and high imaging quality. In recent years, non-standard trajectory scanning and distributed multi-light source imaging have greatly expanded the application scenarios of CT systems, and have shown great potential in solving problems such as large-channel imaging and high-speed imaging. 

Today’s CT machine is a complex system with gantry rotation. Inevitably, the weight of slip ring motor and the centrifugal force severely limits the rotation speed, thus limits the sweep speed of the whole system as well. To overcome this problem, stationary CT system has became a good choice. Avilash Cramer (et.al) designed a stationary computed tomography for space and other resource-constrained environments \cite{bib27}. Tao Zhang(et.al) designed a stationary computed tomography with X-ray source and detector in linear symmetric geometry \cite{bib28}.  Hongguang Cao(et.al) proposed a stationary real-time CT imaging system, comprising an annular photon counting detector, an annular scanning x-ray source, and a scanning sequence controller \cite{bib29}. Derrek Spronk(et.al) designed a stationary CT system with the usage of carbon nanotube for head and breast CT \cite{bib30}\cite{bib31}. However, the energy level is limited so that it can only be applied for human body tissues, when it comes with industrial productions, especially metal objects, X-ray with much higher energy is needed to penetrate the inspected objects. If we use this architecture with traditional X-ray tube, the volume and heat dissipation requirements are difficult to be satisfied.

To maintain the short-time advantage and overcome the spatial limitation of stationary CT, we proposed a brand-new CT system structure named pipelined real-time industrial CT inspection system. A inspection module consists of a pair of X-ray source and detector, 60 modules distribute not only radially, but also along the axis in space to acquire projections up to 60 angles, which is sufficient for high quality CT reconstruction based on the traditional FBP algorithm and deep learning neural network. If the inspection modules distribute only radially, they will interfere with each other, the distribution along the axis can avoid this problem. This structure can be easily integrated to the production line as well.

\section{Hardware design}
\subsection{The system structure}
The main parts of our CT system include fan beam X-ray source, linear X-ray detector. In order to reduce the space needed between the scanner, we consist these two main parts into a pair of detection module. As is shown in figure1(c), the imaging modules in group 1 and in group 2 are staggered distributed, the circumferential distance is 105 degrees.

\begin{figure}[h]
\centering  
\subfigure[linear X-ray detector]{
\label{Fig1.(a)}
\includegraphics[width=0.25\textwidth]{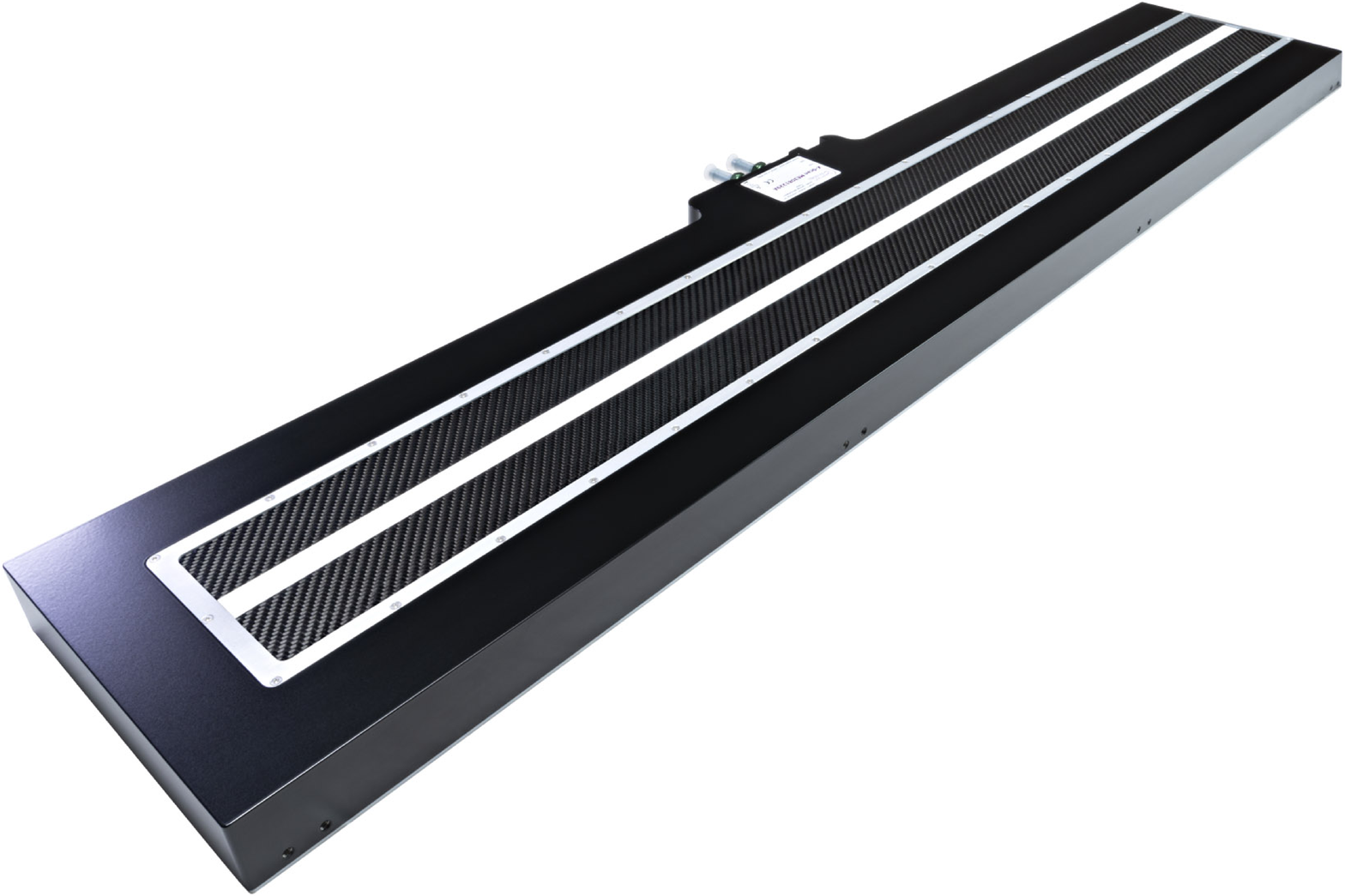}}
\hspace{1.5cm}
\subfigure[integrated fan beam X-ray source]{
\label{Fig1.(b)}
\includegraphics[width=0.25\textwidth]{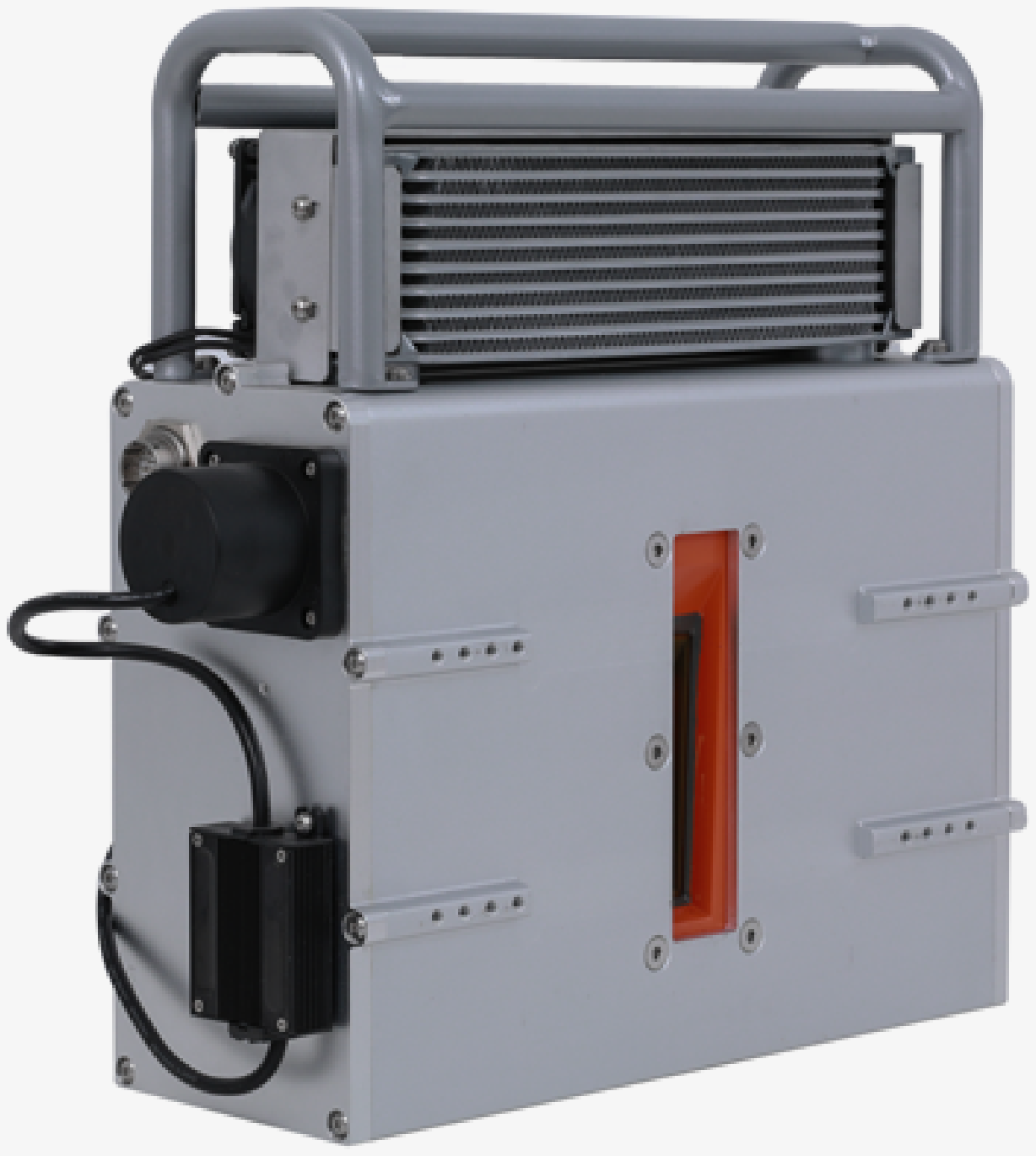}}
\hspace{1.5cm}
\subfigure[two adjacent imaging modules]{
\label{Fig1.(c)}
\includegraphics[width=0.25\textwidth]{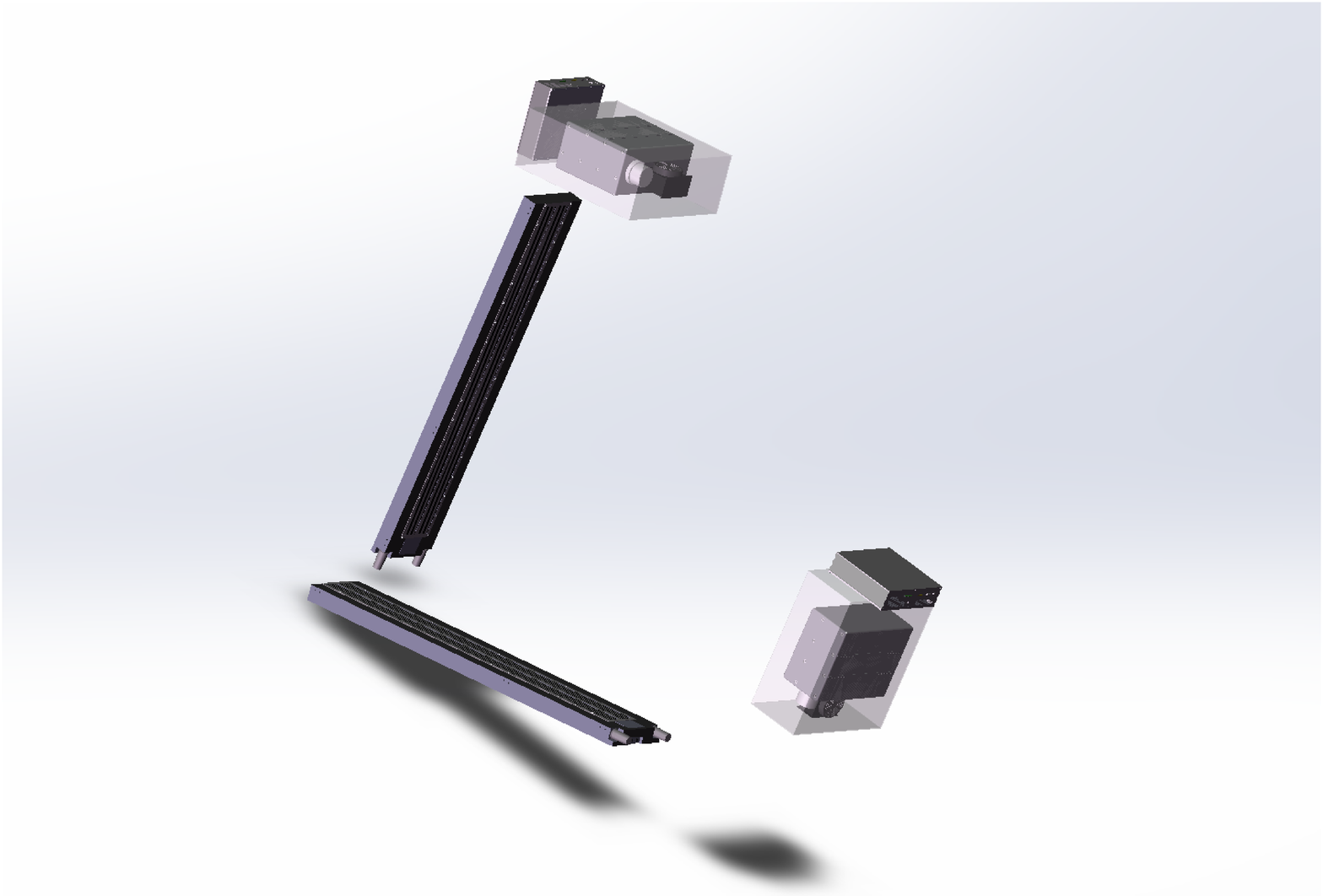}}
\caption{Main parts of the system}
\label{Fig1}
\end{figure}

\begin{table}[h]
\centering\small
\caption{The detector specifications}\label{tab1}%
\begin{tabular}{@{}cc@{}}
\toprule
Feature & Value\\
\midrule
Detection method    & Photon counting\\
Crystal thickness    & 2mm\\
Detector element pitch    & 0.8mm\\
Detector element binning    & 1x1 (0.8mm pitch)\\
Active area length    & 1229 mm (1536 pixels)\\
Energy range    & 20 to 160keV\\
Line speed    & 4m/min to 96m/min\\
Counting period    & 0.5ms to 100ms\\
Dead time    & 10 µs\\
Energy bins (channels)    & Up to 128\\
Pixel dynamic range/ energy bin    & 16 bits per bin\\
\bottomrule
\end{tabular}

\end{table}

The detector specifications of our system were illustrated in Table 1, inspected slice numbers in each object can be calculated based on those parameters:

\begin{equation}
K=\frac{L}{S\cdot (P+T)}
\end{equation}                                                                                                                                
In Eq. (1), K refers to the inspected slice numbers in each object, L(m) refers to object length, S(m/s) refers to line speed, P(s) refers to counting period and T(s) refers dead time. To make sure the inspected slices in different objects are aligned, e.g., location of slice1 in object1 is consistent with location of slice1 in object2. Only in this way will guarantee projections in different angles of each slice locate in the same plane, which avoids distortion during the reconstruction process.
Frame rate for our real-time imaging can be calculated as follows:

\begin{equation}
FPS=\frac{1}{(P+T)}
\end{equation}  

Imaging quality can be modified by controlling the number of energy bins:

\begin{equation}
Q=A\cdot w\cdot b\cdot \Pr
\end{equation}

Where Q refers to the quality of projections, A refers to Active area length, w refers to Detector element width, b refers to energy bins and Pr refers to Pixel dynamic range. Projection quality determines the quality of reconstructed slices. 

The equipment is based on the principle of CT slice imaging, the difference is that the X-ray source and detector through the spiral array arranged with the detected work-piece around the motion direction, for the work-piece that is panning on the conveyor belt, whenever one of the interfaces passes through all the detection modules, the reconstruction process is completed.
In traditional rotation CT machine, the amount of rotation needed is between half revolution and a full revolution: more precisely the minimal rotation angle $\varphi$ is:

\begin{equation}
\varphi =\pi +2\cdot \arcsin (\frac{r}{R})
\end{equation}

where r is the radius of the field of view, and R is the distance between the source and the axis of rotation. One slice of volume is complete when all the needed projections are back projected. In our system, r is around 280mm, and R is around 1661mm. Based on Eq. (4), the minimal rotation angle is 199.41 degrees. We designated the distribution angle of all the detection module as 206.5 degrees for calculation convenience.

\begin{figure}[!h]%
\centering
\includegraphics[width=0.6\textwidth]{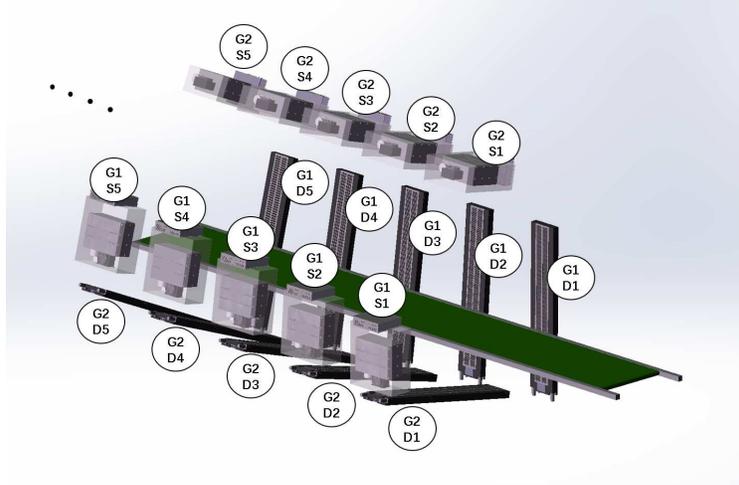}
\caption{Scenograph of the system}\label{Fig2}
\end{figure}

As is seen in figure2, G1S1-G1S5 belong to the first group of the X-ray sources, G1D1-G1D5 belong to the first group of detectors, G2S1-G2S5 belong to the second group of x-ray sources and G2D1-G2D5 belong to the second group of detectors. G stands for group, S stands for source and D stands for detector.

In the reconstruction process, it is necessary to obtain a sufficient angle of projection data in order to ensure the image quality of the reconstructed slices, The device obtains cross-section projection data at 60 different angles for slice image reconstruction through the arrangement of 60 pairs of X-ray sources and detectors.
The circumferential angle interval between each group of light source detectors is 3.5°, and the total detection angle is 206.5°. In order to reduce the minimum distance between each pair of X-ray source and detector, and make the equipment more compact as a whole, a space-interleaved arrangement is adopted.

\begin{figure}[h]
\centering  
\subfigure[overall view of the system]{
\label{Fig3.(a)}
\includegraphics[width=7cm]{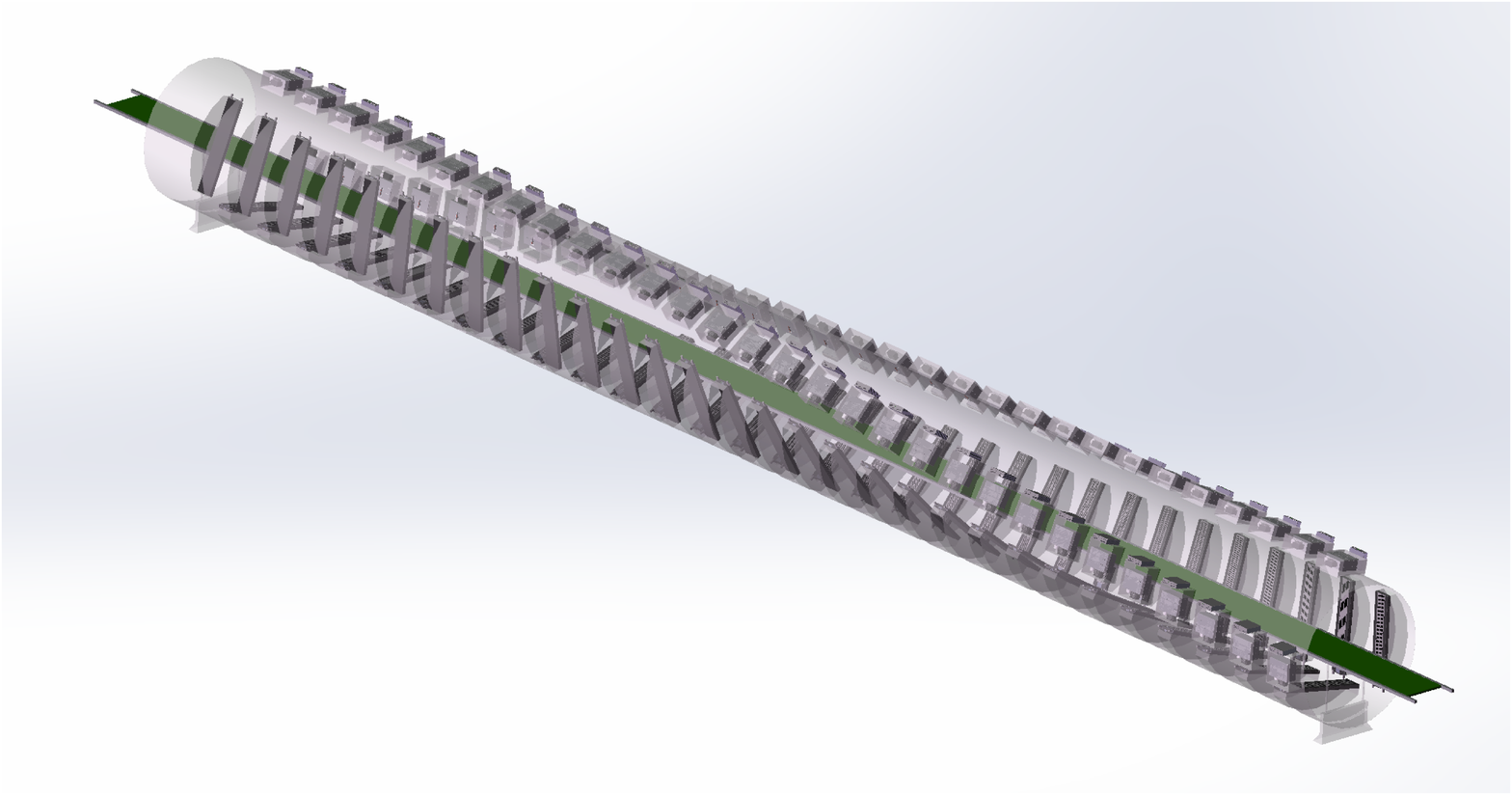}}
\hspace{0.15\textwidth}
\subfigure[side view of the system]{
\label{Fig3.(b)}
\includegraphics[width=5.3cm]{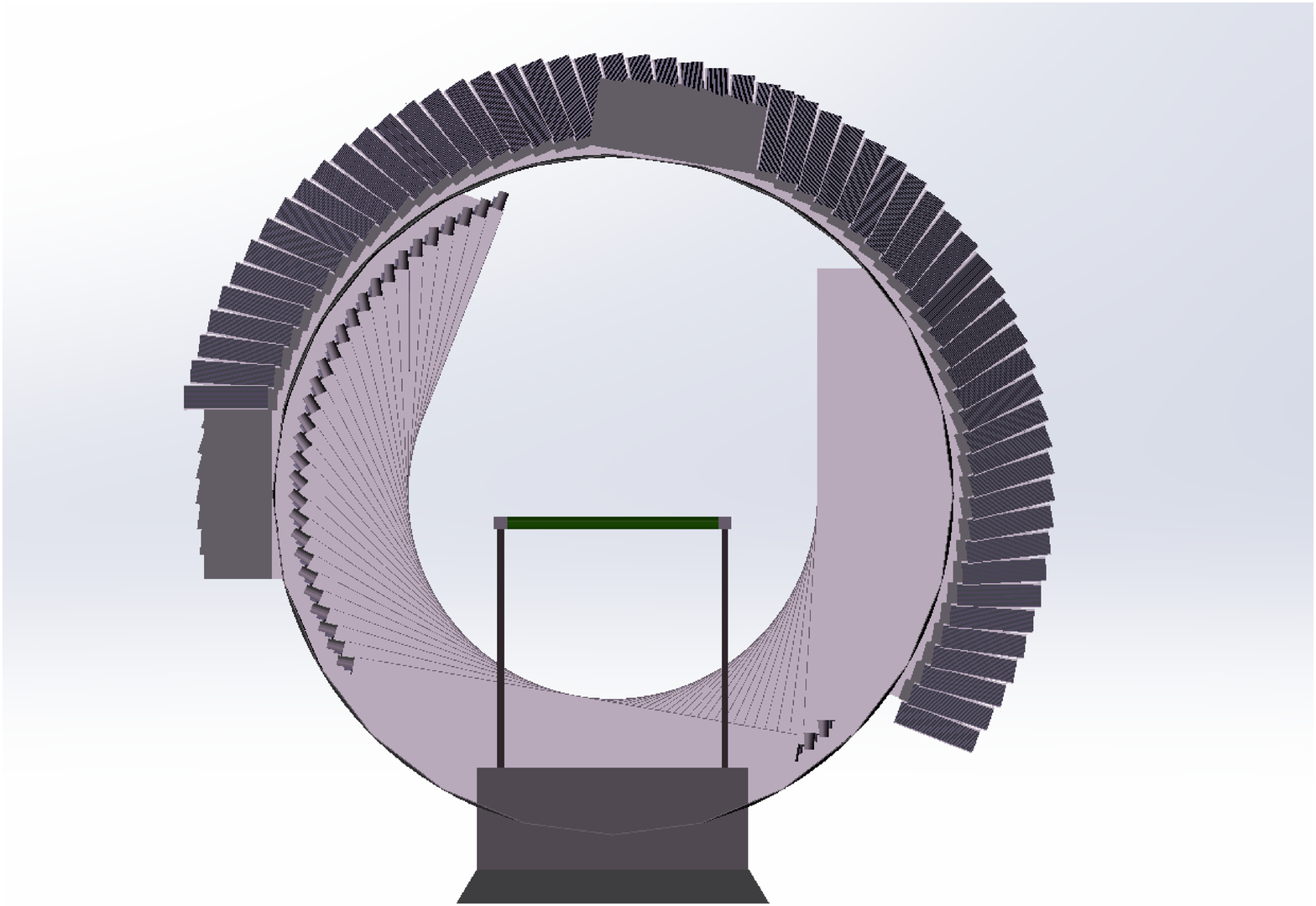}}
\caption{Overall view and side view of the system}
\label{Fig3}
\end{figure}

From 0 ° to 101.5 °, X-ray sources and detectors are arranged every 3.5 °, a total of 30 pairs are arranged, forming the first group. From 105 ° to 206.5 °, other pairs of X-ray sources and detectors are arranged every 3.5 ° for a total of 30 pairs as well, forming a second group, and the two groups add up to a total of 60 pairs of X-ray sources and detectors, the overall view and side view of the system is shown in figure3.

With the development of modern industry, industrial CT plays a major role in non-destructive testing and reverse engineering. The results of non-destructive testing of products using industrial CT show that industrial CT technology has high detection sensitivity for various common defects such as porosity, inclusion, pinhole, shrink hole, and delamination. It can accurately determine the size of these defects and locate their position in the object as well. Compared with other conventional non-destructive testing technologies, the spatial and density resolution of industrial CT technology is less than 0.5\%, the imaging size accuracy is high, and it is not limited by the type and geometry of the work-piece material. It can generate three-dimensional images of material defects, which is of great research and application value in the detection of defects such as structural dimensions, material uniformity, micro-pore rate and overall micro-cracks, inclusions, porosity, and abnormally large grains in the work-piece to be inspected.

\subsection{Data processing}\label{subsec2}
The data collected by the detectors was sent to the cluster workstation with GPU for processing. The X-ray intensity data after attenuation was divided by the X-ray intensity data without any attenuation. Projections were acquired after this process. The data transmission network is shown in figure4.

\begin{figure}[!htb]
\centering
\includegraphics[width=0.5\textwidth]{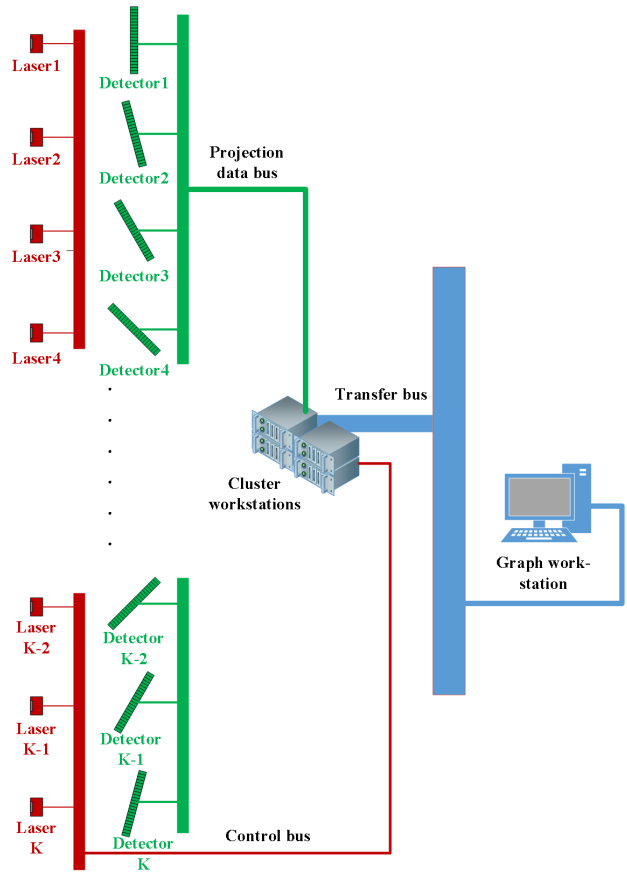}
\caption{Data transmission network}
\label{Fig4}
\end{figure}

Usually, a rotary CT scanner acquires all the projection data after one rotational scan and carry out the reconstruction procedure. Before the projection data of the next slice was collected, FBP reconstruction procedure in GPU has already finished. Therefore, the parallel computing power of GPUs is not well tapped and the low computing efficiency is not suitable for real-time imaging.  In our proposed structure, after projection data in the last angle was acquired by the detector, reconstruction procedure in this slice of interest was executed. By the time when the last projection of next slice of interest was collected, which is much shorter than the imaging algorithm, reconstruction procedure of this new slice of interest was executed parallelly in GPU. Thus, parallel computing power of GPUs is well exerted. Time sequence of data processing is shown in figure5.

\begin{figure}[!htb]%
\centering
\includegraphics[width=0.7\textwidth]{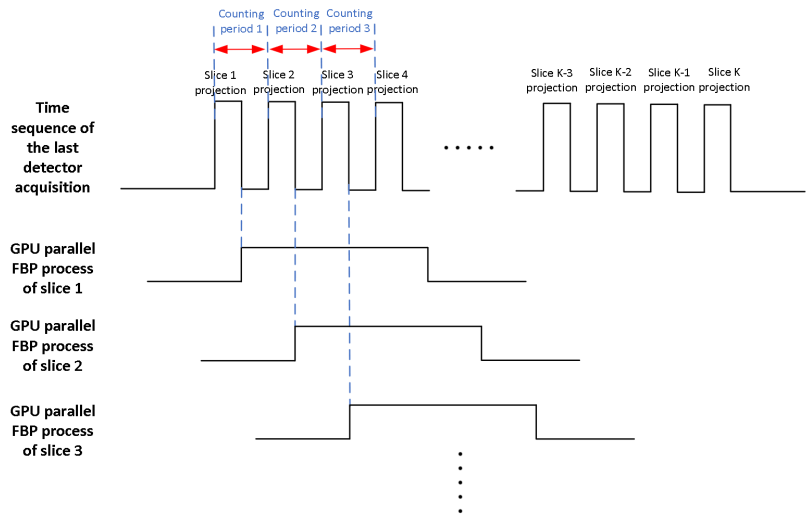}
\caption{Time sequence of data processing}
\label{Fig5}
\end{figure}

In our implementation we preferred to create a continuous reconstruction scheme where the filtered and back-projected slice enter in the system as soon as a new series of projections were read in. In this way, we have a continuous stream of projections that were received from the detectors and a continuous stream of reconstructed slices that was sent to the image processing stage. In such way, objects can be continuously processed even if they only have a small gap between them on the conveyor.
We implemented the FBP algorithm on the cluster workstation with GPU to do the filtered back-projection after the projections were read by detectors. The reconstruction results were then sent to the graphics workstation for post processing by deep-learning.

\section{Reconstruction algorithms}\label{sec4}
\subsection{Filtered back projection}\label{subsec1}
The formation and reconstruction of CT images are mathematically described as the Radon transform and the Radon inverse transform, respectively. To put it simply, the Radon transform is a mathematical representation of the process of forming a CT image by X-ray scanning of an object, while the Radon inverse transformation describes a mathematical method of reconstructing CT projection data and reducing it to an object image. The Radon inverse transformation is actually a two-dimensional Fourier inverse transformation mathematically, its mathematical expression is as follows:

\begin{equation}
    \begin{split}
     f(x,y) & =\int_{-\infty }^{\infty }{\int_{-\infty }^{\infty }{F(u,v){{e}^{j2\pi (ux+vy)}}dudv}} \\ 
     & =\int_{0}^{2\pi }{\int_{0}^{\infty }{F(\omega \cos \theta ,\omega \sin \theta ){{e}^{j2\pi \omega (x\cos \theta +y\sin \theta )}}\omega d\omega d\theta }}
    \end{split}
\end{equation} 

Due to the large amount of calculation and long time consuming of the two-dimensional Fourier inverse transformation, the specific implementation method of the computer for the Radon inverse transformation is mainly the back-projection method, and the back-projection method is divided into direct back-projection method and filtered back-projection method.
Averaging error during the smear, error caused by interpolation and star-like artifacts caused by the overlay of the back projection bring 1/r effect. The following convolutional relationship exists between the refactored image ${{f}_{b}}(x,y)$ and the real original image \textit{f(x,y)}:

\begin{equation}
{{f}_{b}}(x,y)=f(x,y)\cdot \frac{1}{r}   
\end{equation}

In Eq. (6), 1/r is also known as a fuzzy factor, The most commonly used method of removing the 1/r effect is filtered and re-projected method.

As can be seen from Eq. (6) above, 1/r is convolutional with the original image, and after transitioning to the frequency domain by the Fourier transform, it will become a multiplicative relationship, which means that it is easier to process directly at frequency. In the frequency domain, r is called the weight factor, which is actually equivalent to a filter. Method1 only takes three steps to remove the influence of the blur factor and reconstruct the original image:
\begin{itemize}
\item Step1: Perform a two-dimensional Fourier transform on the reconstructed image ${{f}_{b}}(x,y)$ to obtain ${{F}_{b}}(x,y)$;
\item Step2:  Multiply ${{F}_{b}}(x,y)$ by the filter r in the frequency domain;
\item Step3: Take r × ${{F}_{b}}(x,y)$ for a two-dimensional Fourier inverse transformation to obtain the original figure \textit{f(x,y)}.
\end{itemize}

Eq. (5) is the mathematical expression for Method 1.

Method2 of first back-projecting and then filtering requires the use of two two-dimensional Fourier transforms, which needs large amount of computation. This kind of method is not particularly ideal. The filtering and back-projection method, as the name suggests, is a reconstruction method of filtering first and then re-projecting. The specific steps are as follows:
\begin{itemize}
\item Step1: Transform the projection signals at each angle into one-dimensional Fourier;
\item Step2: Filter all projected signals in the frequency domain, that is, multiplied by the weight factor r;
\item Step3: Transform inversely all filtered signals in one dimension and restored to the time domain;
\item Step4: back-project and finally superimpose each filtered projection signal.
\end{itemize}

The advantage of this method is that the two-dimensional Fourier transforms are changed into two one-dimensional Fourier transforms, and the calculation speed is greatly improved. One-dimensional Fourier transform of the projection of $\rho$ is:

\begin{equation}
G(\omega ,\theta )=\int_{-\infty }^{\infty }{g(\rho ,\theta ){{e}^{-j2\pi \omega \rho }}d\rho }
\end{equation}

The mathematical expression for Method 2 is:

\begin{equation}
    f(x,y)=\int_{0}^{2\pi }{{{[\int_{-\infty }^{\infty }{|\omega | G(\omega ,\theta ){{e}^{j2\pi \omega \rho }}}d\omega ]}_{\rho =x\cos \theta +y\sin \theta }}}d\theta 
\end{equation}

The core problem of filtering and in-projection becomes how to choose a suitable filter r. Filter r in mathematics is an idealized filter function that does not exist in reality, so only an approximate filter function can be designed to replace the role of filter r. R-L filter function and S-L filter function are commonly used filter functions. To eliminate the ringing effect caused by window function, smoothing window function is used:

\begin{equation}
    h(\omega )=\left\{ \begin{matrix}
   c+(c-1)\cos \frac{2\pi \omega }{M}, & 0\le \omega \le (M-1)  \\
   0, & other  
\end{matrix} \right.
\end{equation}

This slope filter function is called Hamming window when c = 0.54 and Hann window when c = 0.5 \cite{bib32}.\\
Basically, filtered back projection (FBP) is a convolution operation on the basis of the back projection. The specific process of using filter inversion to reconstruct the image is to first convolve the original data obtained by the detector with a filter function to obtain the projection function of convolution in each direction; and then to reverse project them from all directions, that is, to distribute them evenly to each matrix element according to its original path, and to obtain the CT value of each matrix element after superposition; and then after appropriate process, the tomographic image of the scanned object can be obtained, and the convolutional projection can eliminate the edge sharpness effect caused by the simple back projection. Filter function compensates for the high-frequency components in the projection and reduces the density of the projection center, and ensures that the edges of the reconstructed image are clear and the internal distribution is uniform. As the number of projections increases, the quality of the reconstruction is increased significantly. We used an open-access Python package Astra-toolbox to operate the FBP process \cite{bib33}. It can be seen from figure 6 that the effect on reconstruction quality with the increasement of projections number.

\begin{figure}[h]
    \centering
    \includegraphics[width=0.9\textwidth]{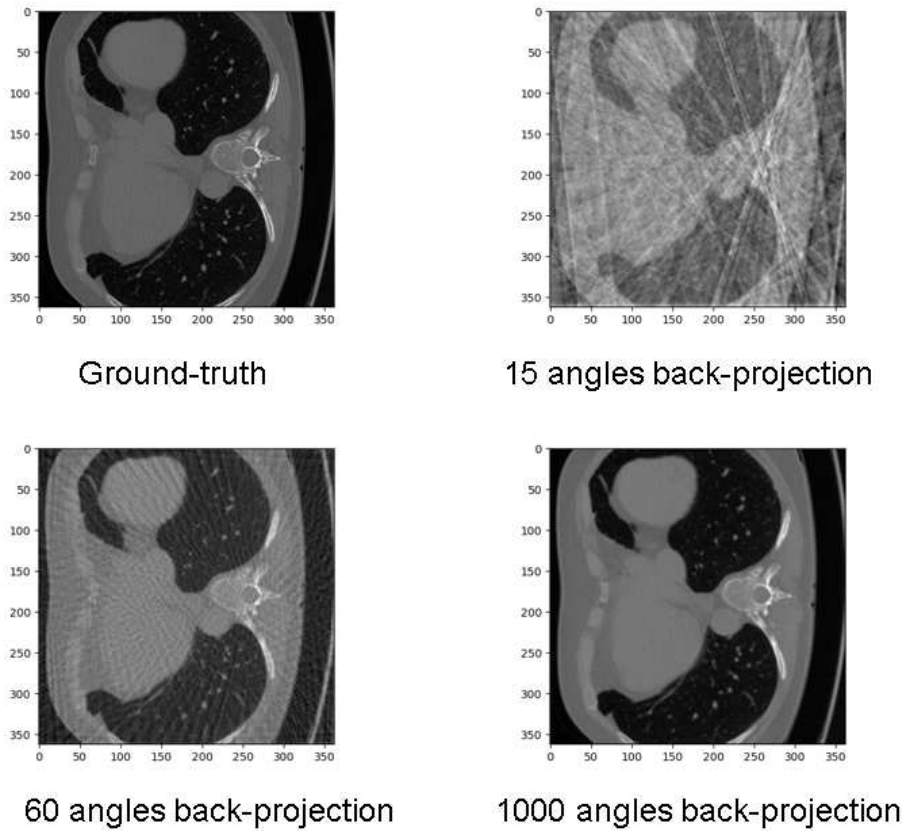}
    \caption{FBP results with different projection numbers}
    \label{Fig6}
\end{figure}

\subsection{Deep convolutional neural network design}\label{sec2}
\subsubsection{Network structure}\label{sec1}
There is no doubt that the CT reconstruction achieves ideal results using FBP algorithm with adequate projection data. However, for the structure of the system we designed, it will be too redundant and complex to use so many pairs of X-ray sources and detectors to acquire adequate projections. Our device will take up too much space and it will be too expensive to build the system. To make our design practical, it is essential to reduce the number of X-ray sources and detectors. Consequently, the CT reconstruction process becomes an ill-posed question, trying to obtain the solution with insufficient parameters. The FBP reconstruction result contains many artifacts especially streaking artifacts, it is obvious that image with those artifacts cannot meet the requirements of industrial inspection. Typically, some post-processing approaches such as denoising were applied to reduce those artifacts. Recent works \cite{bib34}\cite{bib35}\cite{bib36} have successfully used convolutional neural networks such as U-net \cite{bib37} to solve this kind of problem. Basically, we were trying to train an end-to-end neural network to acquire clean reconstruction results out of the noisy FBP reconstruction results from sparse-view projection data. In our design, the sparse-view projection data was acquired through 60 angles which were evenly distributed within 180 degrees. The structure of our sparse-view reconstruction network is shown in figure7.

\begin{figure}[!h]
    \centering
    \includegraphics[width=0.8\textwidth]{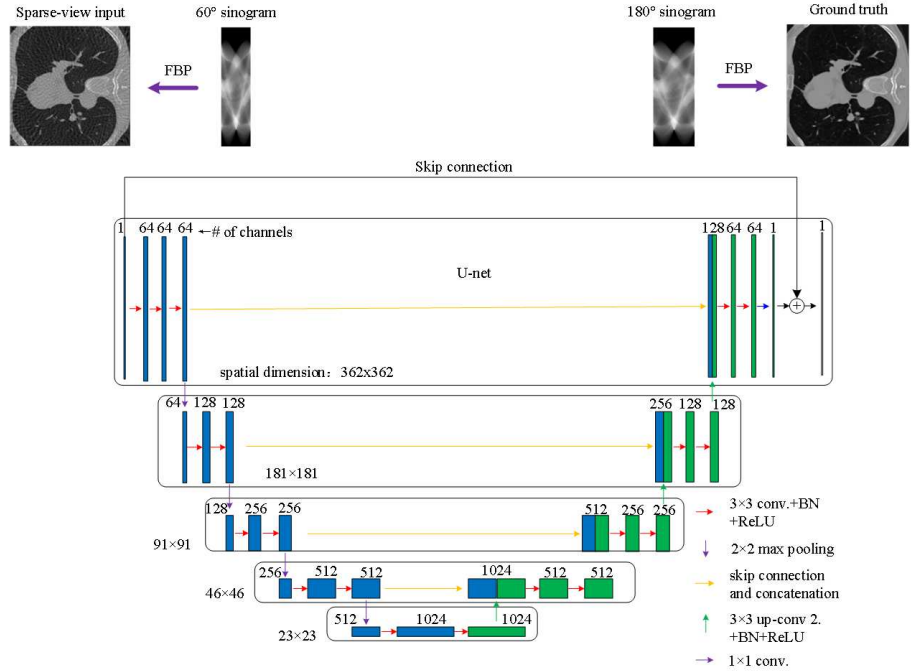}
    \caption{U-net structure}
    \label{Fig7}
\end{figure}

Several details of the network are as follows:
\begin{enumerate}[(1)]
    \item Down-sampling: U-net extracts the features of the input image through max-pooling and two-dimensional convolution operations. In the process of down-sampling, the size of the image feature map continues to shrink, and the number of feature channels continues to increase, which means that more local details of the image are extracted.\\

    \item Up-sampling: De-convolution, more precisely transpose convolution operations were used to gradually restore ground-truth from features. It can be understood as an inverse process of down-sampling, during this process, image features are gradually restored. In order to ensure that the feature scale change during the up-sampling process is symmetrical with the down-sampling process, the kernel size and padding of transposed convolution are variable.\\

    \item Feature fusion: Before every up-sampling process, input features are fused with the features in feature extraction part at the same feature scale. Using the abstract features encoded earlier to restore to the original size, U-net employs a completely different way of feature fusion: concatenation, U-net concatenates features together in the channel dimension to form thicker features. When recovering the down-sampling data, the feature scale will change, there will inevitably be information loss. At this time, the role of feature fusion is highlighted, feature fusion plays a role in supplementing information. U-Net integrates features of different scales, while the jump connection ensures that the features recovered from the up-sampling will not be very rough. Some research found that jump connection can make the network minimizer flatter by visualizing the lose landscape, resulting in less sensitivity to new data and stronger generalization capabilities \cite{bib38}.\\

    \item Residual Learning: Inspired by the good generalization performance of residual network \cite{bib39}, we add a skip connection between input image and output image. Pixels of the input and output image are added linearly, which means that the network essentially learns the difference between input image and output image. The vanishing gradient problem during training can be effectively avoided with this approach.
\end{enumerate}

Our model was designed in Python using PyTorch framework. All the experiments run on Linux system with 24G NVIDIA RTX3090 GPU, Xeon Platinum 8157 CPU @ 3GHz and 86G RAM.
The specifications of our proposed network are shown in the following table2.

\begin{table}[!h]
\renewcommand\arraystretch{0.9}
\centering
\caption{Specifications of the network}\label{tab2}%
\begin{tabular}{@{}cccccc@{}}
\toprule
layer & \begin{tabular}[c]{@{}l@{}}Input \\ channels\end{tabular} & \begin{tabular}[c]{@{}l@{}}Output \\ channels\end{tabular} & \begin{tabular}[c]{@{}l@{}}Kernel   \\ size\end{tabular} & stride & padding \\
\midrule
Conv0 & 1 & 64 & 3 & 1 & ‘same’ \\
Conv1 & 64 & 64 & 3 & 1 & ‘same’ \\
Conv2 & 64 & 64 & 3 & 1 & ‘same’ \\
Max pooling0 & 64 & 64 & 2 & 2 & 0 \\
Conv3 & 64 & 128 & 3 & 1 & ‘same’ \\
Conv4 & 128 & 128 & 3 & 1 & ‘same’ \\
Max pooling1 & 128 & 128 & 2 & 2 & 1 \\
Conv5 & 128 & 256 & 3 & 1 & ‘same’ \\
Conv6 & 256 & 256 & 3 & 1 & ‘same’ \\
Max pooling2 & 256 & 256 & 2 & 2 & 1 \\
Conv7 & 256 & 512 & 3 & 1 & ‘same’ \\
Conv8 & 512 & 512 & 3 & 1 & ‘same’ \\
Max pooling3 & 512 & 512 & 2 & 2 & 0 \\
Conv9 & 512 & 1024 & 3 & 1 & ‘same’ \\
Conv10 & 1024 & 1024 & 3 & 1 & ‘same’ \\
Conv\_trans0 & 1024 & 512 & 2 & 2 & 0 \\
Conv11 & 1024 & 512 & 3 & 1 & ‘same’ \\
Conv12 & 512 & 512 & 3 & 1 & ‘same’ \\
Conv\_trans1 & 512 & 256 & 3 & 2 & 1 \\
Conv13 & 512 & 256 & 3 & 1 & ‘same’ \\
Conv14 & 256 & 256 & 3 & 1 & ‘same’ \\
Conv\_trans2 & 256 & 128 & 3 & 2 & 1 \\
Conv15 & 256 & 128 & 3 & 1 & ‘same’ \\
Conv16 & 128 & 128 & 3 & 1 & ‘same’ \\
Conv\_trans3 & 128 & 64 & 2 & 2 & 0 \\
Conv17 & 128 & 64 & 3 & 1 & ‘same’ \\
Conv18 & 64 & 64 & 3 & 1 & ‘same’ \\
Conv19 & 64 & 1 & 3 & 1 & ‘same’\\
\bottomrule
\end{tabular}
\end{table}

\subsubsection{Dataset}\label{sec2}
The dataset comes from LoDoPaB-CT \cite{bib40}. This dataset crops all included images to consistently 362 × 362 pixels. Consequently, the number of weight parameters in our proposed is greatly reduced and the network is easier to be trained.
One of the big challenges of training a network with good performance is acquiring high quality dataset for training. Unfortunately, industrial CT datasets are usually not available given that patent protection is involved. In this case, we try to train the network using open-source medical datasets and assess the performance using the industrial CT image acquired in our laboratory. Several tricks have been applied to improve the generalization of the model, including data augmentation and L2 regularization.

Data augmentation is effective to improve the desired invariance and robustness properties of the network. Since we want to train the reconstruction network using medical dataset and apply to industrial images, data augmentation can be a good trick. We randomly flip the input images and the corresponding target images, either in X or Y direction, or in both directions.

The model that uses L2 regularization is called ridge regression. loss function in this model would be:

\begin{equation}
    \min \frac{1}{2m}{{\sum\nolimits_{i=1}^{m}{\left( f(x)-{{y}^{(i)}} \right)}}^{2}}+\lambda \left\| w \right\|_{2}^{2}
\end{equation}

Where $\lambda $ is the weight coefficient of the L2 regularization, The larger the $\lambda $, the stronger the restriction on the weight vector. In the process of the L2 regularization, it is generally inclined to make the weights in the network as small as possible. A model with small parameter values is relatively simple and has great adaption to different datasets. L2 regularization avoids the phenomenon of overfitting to a certain extent. It can be imagined that for a linear regression equation, if the parameters are large, then as long as the data is offset a little, it will have a great impact on the result; but if the parameters are small enough, the data offset a little more will not affect the result, and the anti-interference ability is strong.

\subsection{Reconstruction quality assessment}\label{sec3}

\begin{figure}[h]
\centering  
\subfigure[Train loss]{
\label{Fig8(a)}
\includegraphics[width=6cm]{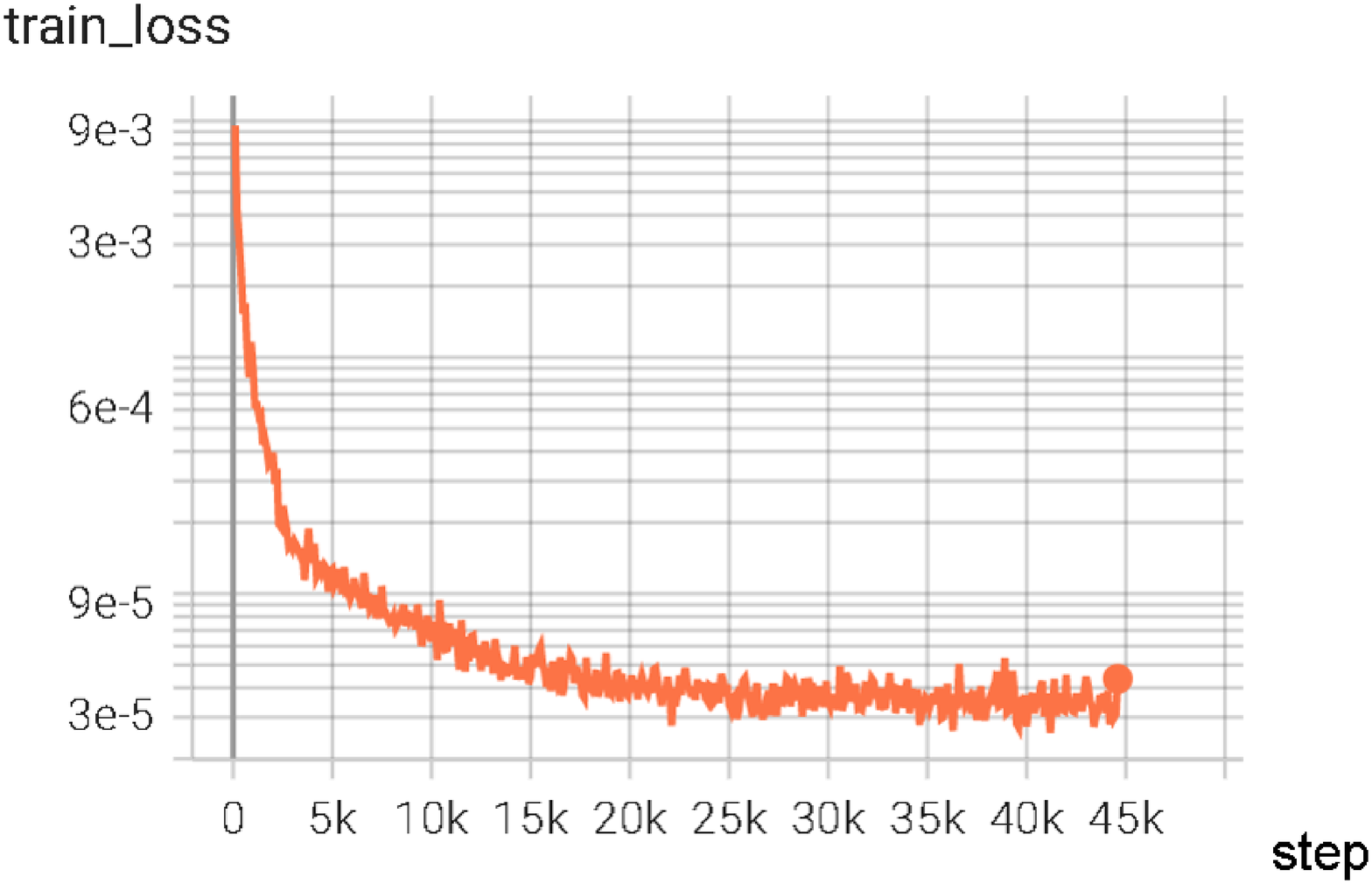}}
\hspace{2cm}
\subfigure[Test loss]{
\label{Fig8(b)}
\includegraphics[width=6cm]{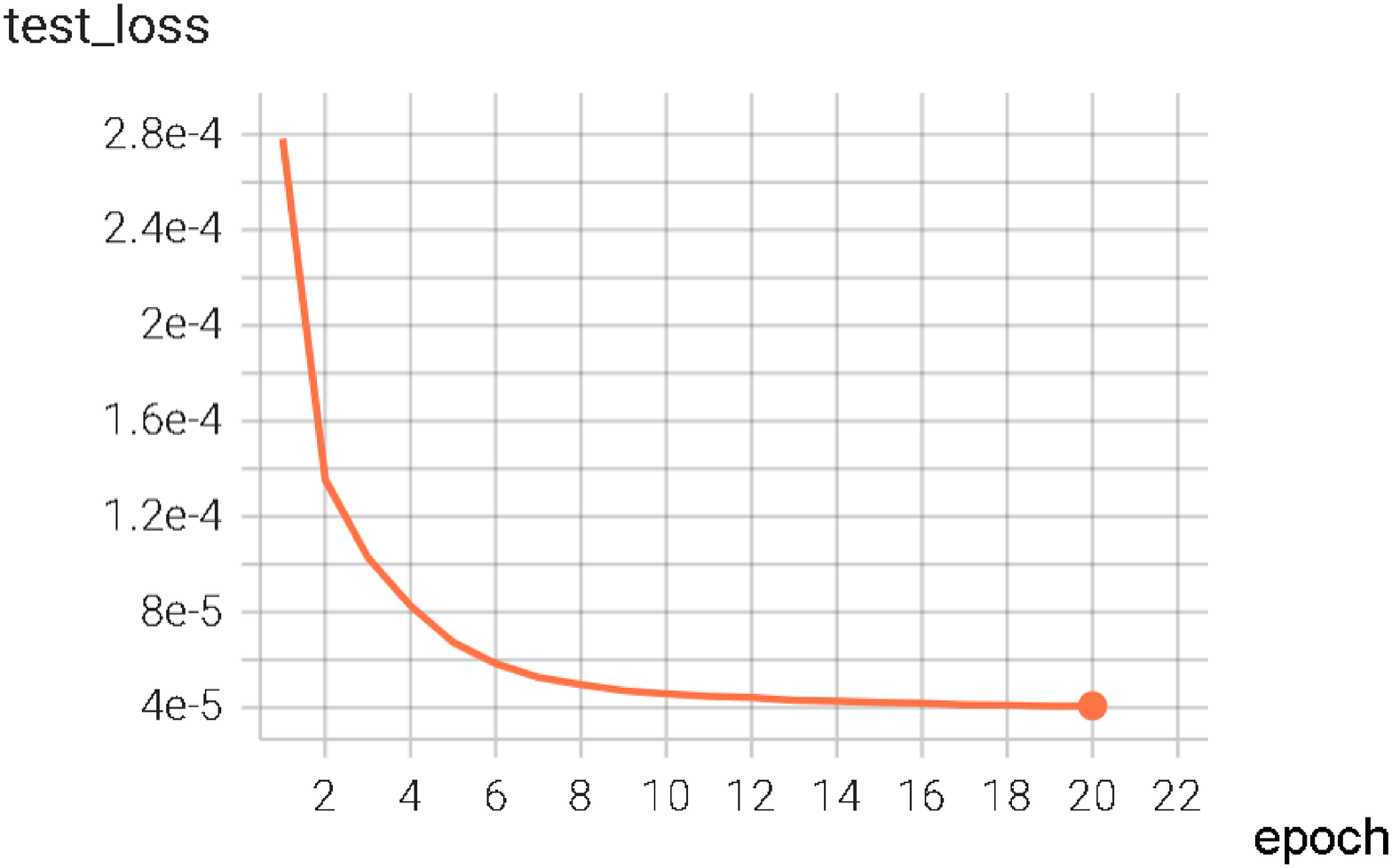}}
\caption{Train loss and test loss}
\label{Fig8}
\end{figure}

In this initial study, for the FBP, Hann filter were used as we mentioned before. For the post-processing approach (FBP + U-Net), we used a U-Net architecture with 5 scales. We trained it using the proposed dataset by minimizing the mean squared error loss with the Adam algorithm for a maximum of 20 epochs with batch size 16. Additionally, we used a learning rate of 1e-5 and the weight parameter of L2 regularization is 1e-7. Training loss and test loss during the training are shown in Figure 8. The model with the highest mean peak signal-to-noise ratio (PSNR) on the test set was selected during training. Reconstructed samples are shown in Figure 9. Region of interest (ROI) of sample 1 is show in figure10, we can see that artifacts caused by under sampling were clearly reduced.

\begin{figure}[h]
    \centering
    \includegraphics[width=1\textwidth]{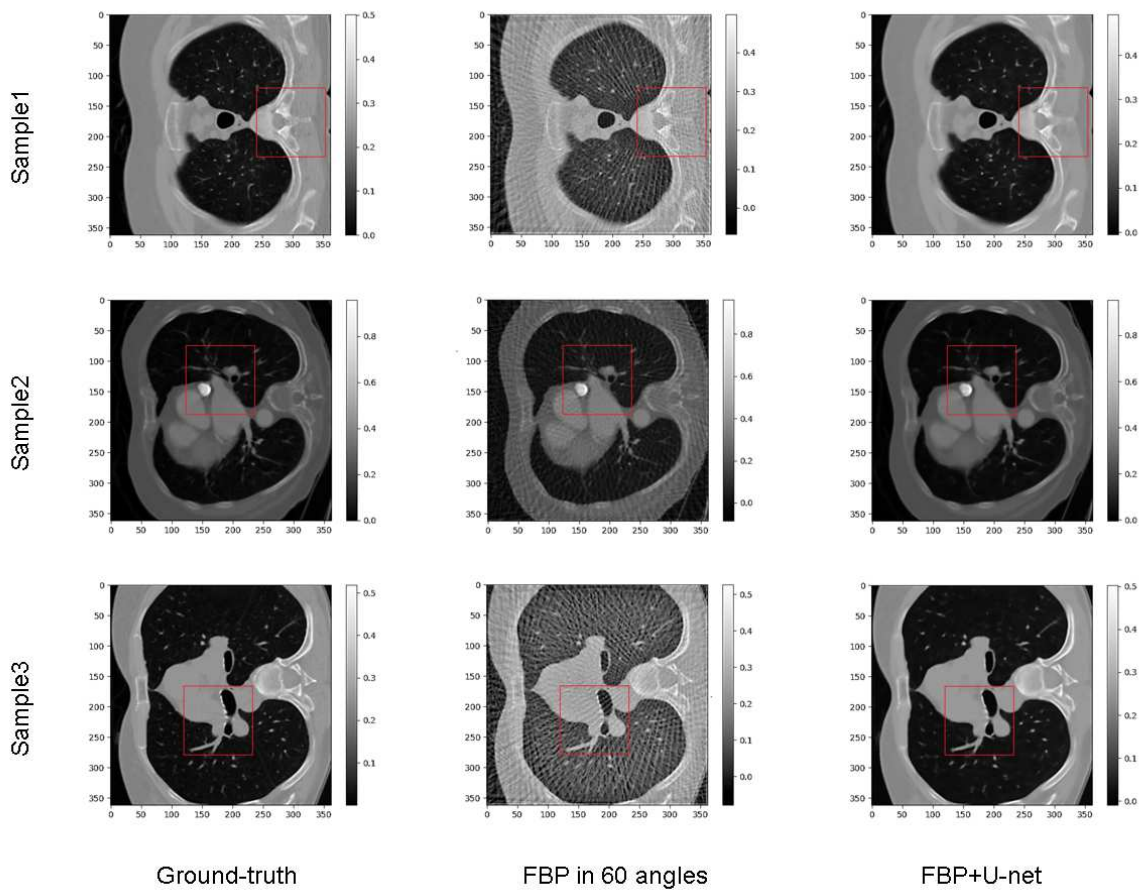}
    \caption{Dataset samples reconstruction}
    \label{Fig9}
\end{figure}

\begin{figure}[h]
    \centering
    \includegraphics[width=0.8\textwidth]{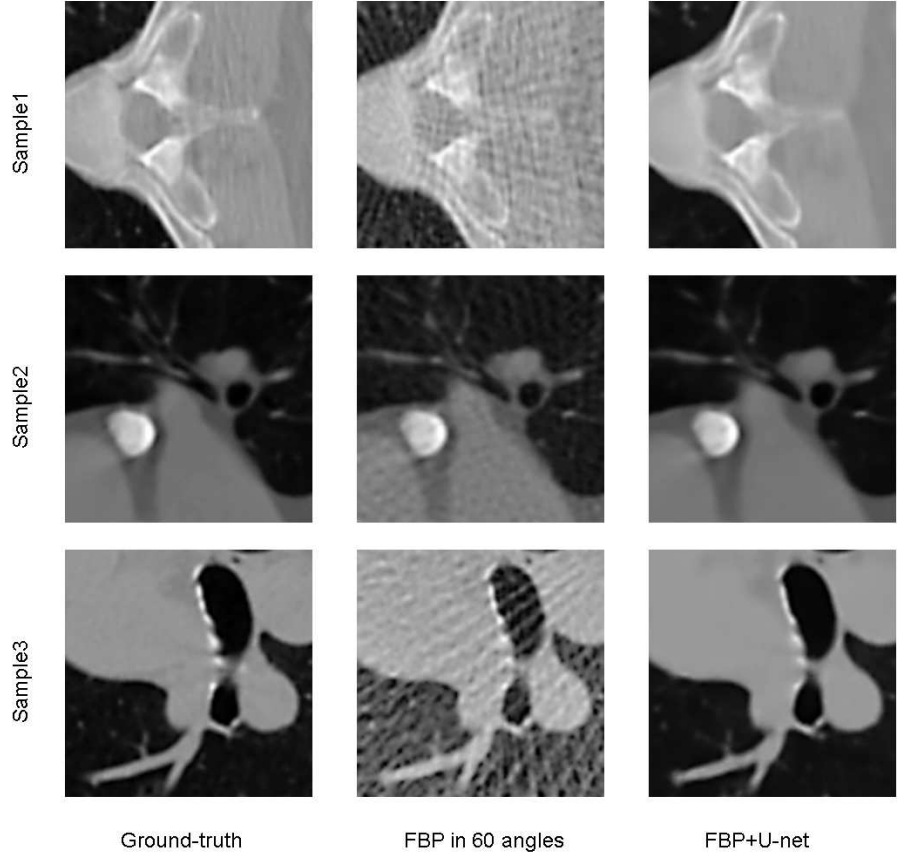}
    \caption{Comparison in region of interest}
    \label{Fig10}
\end{figure}

To validate the reconstruction quality of our proposed deep learning approach, we provide reference reconstructions and quantitative results for the standard filtered back-projection (FBP) and a learned post-processing method (FBP + U-Net).

We use the PSNR to evaluate the reconstruction quality. Peak signal-to-noise ratio (PSNR) is an engineering term that expresses the ratio of the maximum possible power of a signal to the destructive noise power that affects the accuracy of its representation. Since many signals have a very wide dynamic range, the PSNR is often expressed in logarithmic decibel units. It can be written as follows:

\begin{equation}
    MSE=\frac{1}{mn}{{\sum\nolimits_{i=1}^{m-1}{\sum\nolimits_{j=0}^{n-1}{\left[ I(i,j)-K(i,j) \right]}}}^{2}}
\end{equation}

\begin{equation}
    PSNR=10\cdot {{\log }_{10}}(\frac{MAX_{I}^{2}}{MSE})=20\cdot {{\log }_{10}}(\frac{MA{{X}_{I}}}{\sqrt{MSE}})
\end{equation}

$MA{{X}_{I}}$ is the maximum numeric value that represents the color of an image point. As we can see from the formula, the smaller MSE, the larger PSNR, the better image quality.

Structural similarity(SSIM) is used to calculate the similarity of two input images. One of them is a ground truth, the other is a reconstructed image, and SSIM can be used as a measure of quality. The mathematical definition of SSIM is:

\begin{equation}
\begin{aligned}
    SSIM={{\left| l(x,y) \right|}^{\alpha }}{{\left| c(x,y) \right|}^{\beta }}{{\left| s(x,y) \right|}^{\alpha \gamma }}\\
\end{aligned}
\end{equation}

In Eq. (13):

\begin{equation}
    l(x,y)=\frac{2{{\mu }_{x}}{{\mu }_{y}}+{{c}_{1}}}{\mu _{x}^{2}+\mu _{y}^{2}+{{c}_{1}}}
\end{equation}

\begin{equation}
    c(x,y)=\frac{2{{\sigma }_{xy}}+{{c}_{2}}}{\sigma _{x}^{2}+\sigma _{y}^{2}+{{c}_{2}}}
\end{equation}

\begin{equation}
    s(x,y)=\frac{{{\sigma }_{xy}}+{{c}_{3}}}{{{\sigma }_{x}}{{\sigma }_{y}}+{{c}_{3}}}
\end{equation}

l(x,y) is the brightness component, c(x,y) is a contrast component,  s(x,y) is the structural component. ${{\mu }_{\text{x}}}$  and ${{\mu }_{\text{y}}}$ represent the means of x, y, ${{\sigma }_{x}}$  and ${{\sigma }_{y}}$ represent the standard deviation of x, y, respectively. ${{\sigma }_{xy}}$ represents the covariance of x and y.

Table 3 depicts the obtained results in terms of the MSE, PSNR and SSIM metrics during the training compared with the FBP results. 100 samples in test dataset were used to calculate those metrics. As it can be observed, in the process of MSE reduction, the FBP+U-net approach outperforms the classical FBP reconstructions by a margin of 10 dB in PSNR and the SSIM improved by 38\%. This demonstrates that our proposed reconstruction method achieves good performance in test dataset.

\begin{table}[h]
\centering
\caption{Reconstruction metrics for test dataset }\label{tab3}%
\begin{tabular}{@{}ccccccc@{}}
\toprule
 & \begin{tabular}[c]{@{}l@{}}FBP in 60 \\ angles\end{tabular} & \begin{tabular}[c]{@{}l@{}}Epoch \\ 1\end{tabular} & \begin{tabular}[c]{@{}l@{}}Epoch\\5 \end{tabular} & \begin{tabular}[c]{@{}l@{}}Epoch\\10 \end{tabular} & \begin{tabular}[c]{@{}l@{}}Epoch\\15 \end{tabular} & \begin{tabular}[c]{@{}l@{}}Epoch\\20 \end{tabular}\\
\midrule
MSE & \begin{tabular}[c]{@{}l@{}}4.13e-4$\pm $\\2.87e-8\end{tabular} 
& \begin{tabular}[c]{@{}l@{}}2.58e-4$\pm $\\2.06e-8\end{tabular} 
& \begin{tabular}[c]{@{}l@{}}6.96e-5$\pm $\\ 2.41e-9\end{tabular} 
& \begin{tabular}[c]{@{}l@{}}4.79e-5$\pm $\\1.72e-9\end{tabular} 
& \begin{tabular}[c]{@{}l@{}}4.44e-5$\pm $\\ 1.62e-9\end{tabular} 
& \begin{tabular}[c]{@{}l@{}}4.29e-5$\pm $\\1.68e-9 \end{tabular}\\
\hline
PSNR & \begin{tabular}[c]{@{}l@{}}29.22$\pm $\\ 4.87dB\end{tabular} 
& \begin{tabular}[c]{@{}l@{}}31.39$\pm $\\4.46dB\end{tabular} 
& \begin{tabular}[c]{@{}l@{}}37.23$\pm $\\ 6.10dB\end{tabular} & \begin{tabular}[c]{@{}l@{}}39.03$\pm $\\ 7.48dB\end{tabular} & \begin{tabular}[c]{@{}l@{}}39.40$\pm $\\ 7.85dB\end{tabular} & \begin{tabular}[c]{@{}l@{}}39.59$\pm $\\ 7.99\end{tabular} \\
\hline
SSIM & \begin{tabular}[c]{@{}l@{}}0.69$\pm $\\3.90e-3\end{tabular} 
& \begin{tabular}[c]{@{}l@{}}0.79$\pm $\\4.56e-3\end{tabular} 
& \begin{tabular}[c]{@{}l@{}}0.92$\pm $\\ 1.60e-3\end{tabular} & \begin{tabular}[c]{@{}l@{}}0.94$\pm $\\ 1.46e-3\end{tabular} & \begin{tabular}[c]{@{}l@{}}0.94$\pm $\\ 1.34e-3\end{tabular} & \begin{tabular}[c]{@{}l@{}}0.95$\pm $\\  1.44e-3\end{tabular}\\
\bottomrule
\end{tabular}
\end{table}

To verify the generalization of the network on industrial CT, we used CT images of spectral interferometer for verification, which was acquired by the Shimadzu CT instrument at the tube voltage of 200kV and the tube current of 70 ${{\mu }_{A}}$. Figure 11 shows the reconstruction comparison of industrial CT. Table 4 shows that the model also has great performance on industrial data. We recorded the change of different metrics in both test dataset and industrial data during the training process in Figure 12.

\begin{figure}[h]
\centering  
\subfigure[Ground-truth]{
\label{Fig11(a)}
\includegraphics[width=5cm]{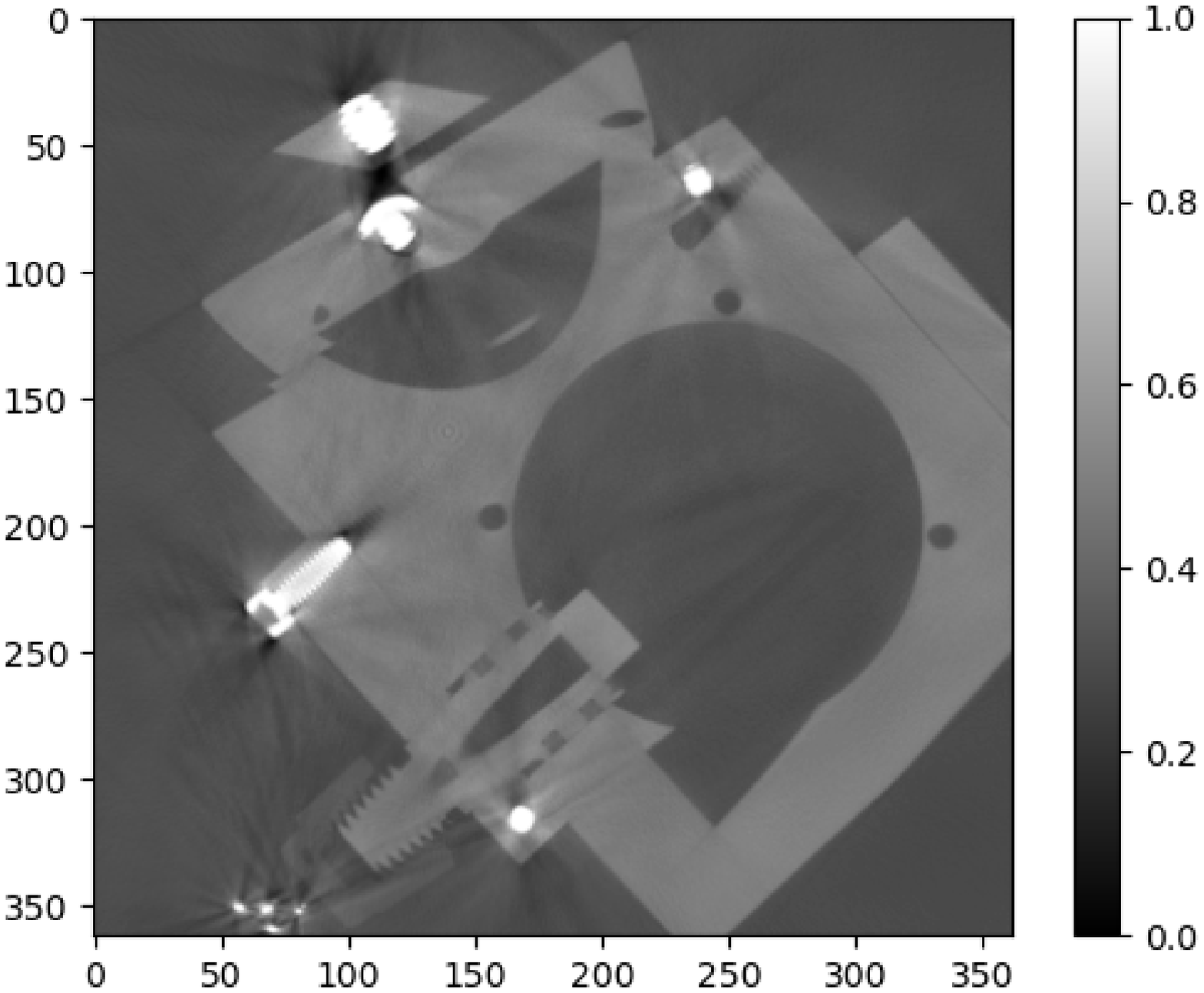}}
\subfigure[FBP in 60 angles]{
\label{Fig11(b)}
\includegraphics[width=5cm]{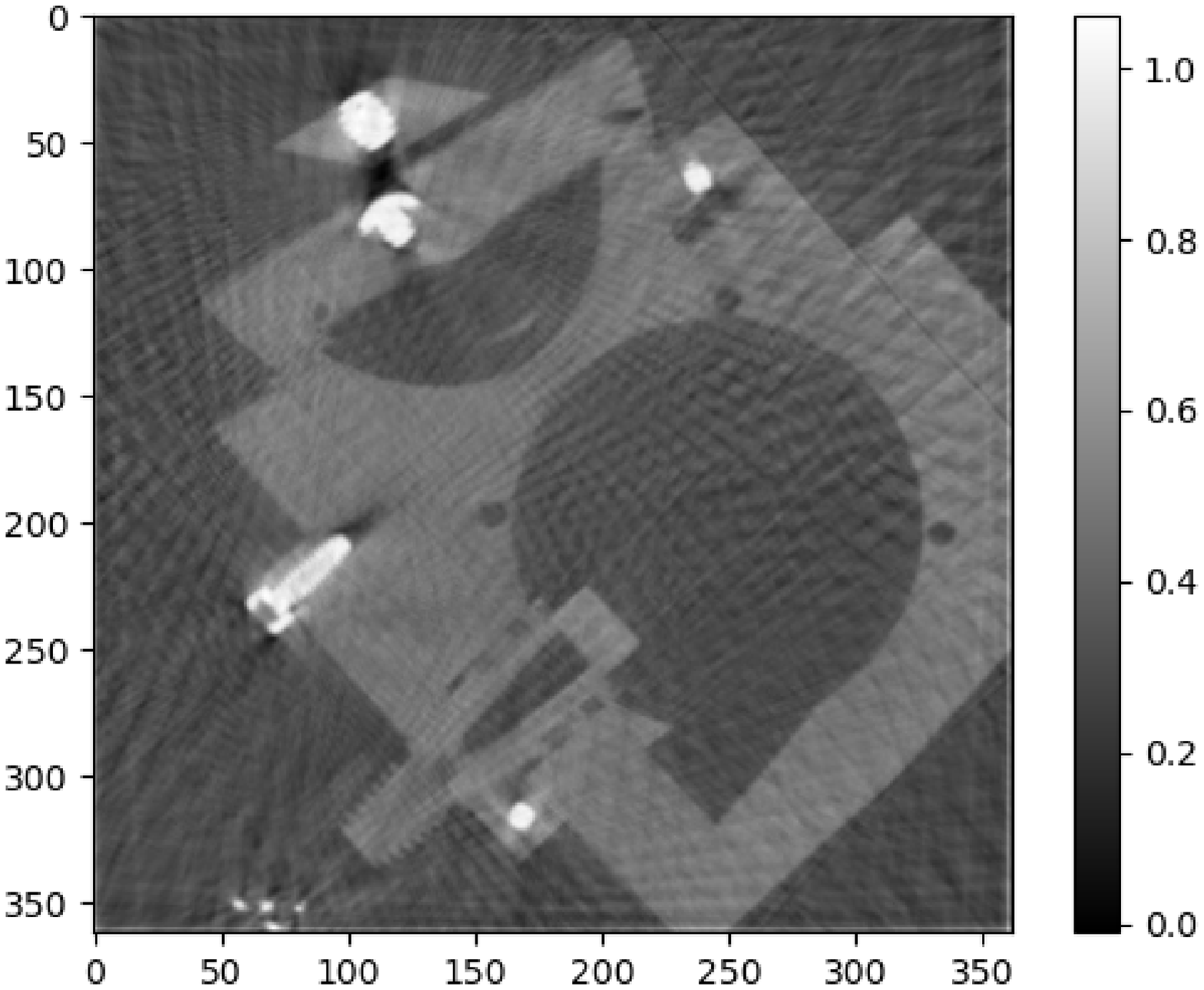}}
\subfigure[FBP+U-net]{
\label{Fig11(c)}
\includegraphics[width=5cm]{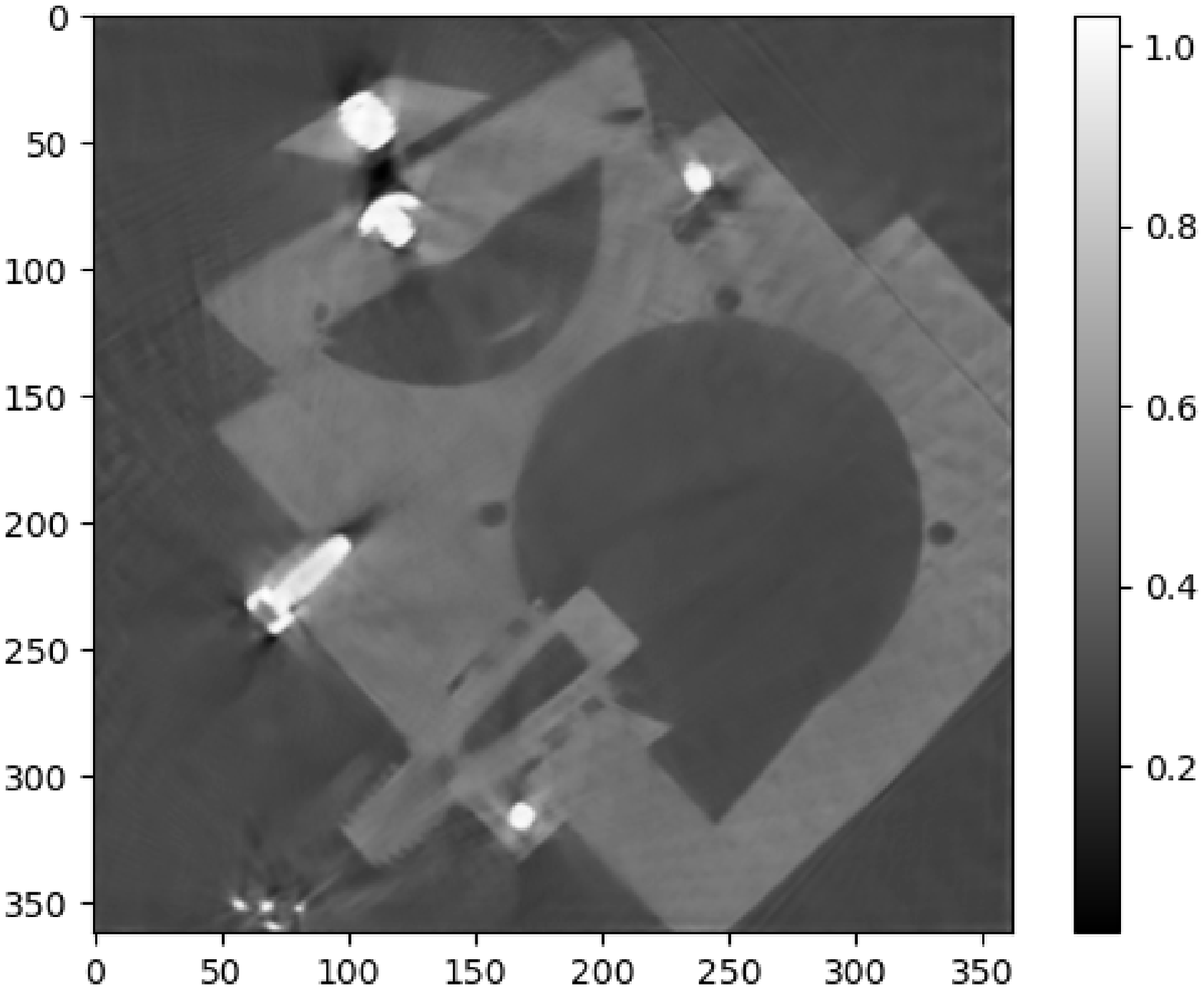}}
\caption {Industrial data reconstruction}
\label{Fig11}
\end{figure}

\begin{figure}[h]
\centering  
\subfigure[MSE]{
\label{Fig12(a)}
\includegraphics[width=8cm]{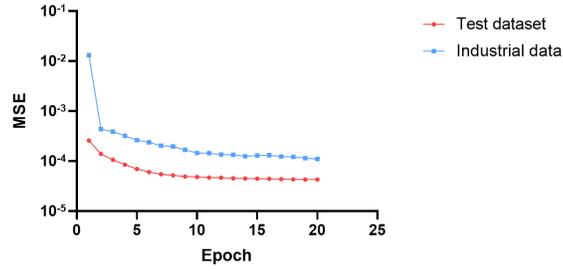}}
\\
\subfigure[PSNR]{
\label{Fig12(b)}
\includegraphics[width=8cm]{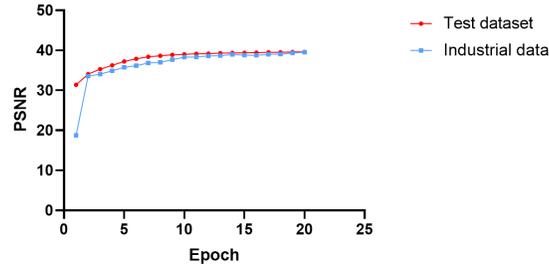}}
\\
\subfigure[SSIM]{
\label{Fig12(c)}
\includegraphics[width=8cm]{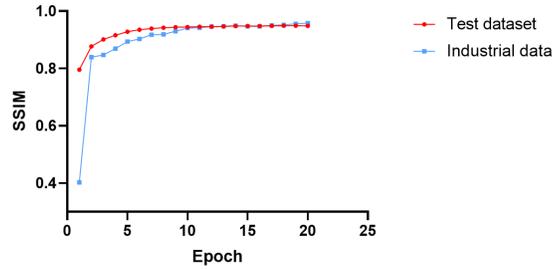}}
\\
\caption {Metrics changes in training process}
\label{Fig12}
\end{figure}

\begin{table}[h]
\centering
\caption{Reconstruction metrics for industrial data }\label{tab4}%
\begin{tabular}{@{}ccc@{}}
\toprule
 & \begin{tabular}[c]{@{}l@{}}FBP in 60 angles\end{tabular}
 & \begin{tabular}[c]{@{}l@{}}FBP+U-net\end{tabular}\\
\hline
MSE & \begin{tabular}[c]{@{}l@{}}7.75e-4\end{tabular} 
& \begin{tabular}[c]{@{}l@{}}1.09e-4\end{tabular}\\ 
\hline
PSNR & \begin{tabular}[c]{@{}l@{}}31.04dB\end{tabular} 
& \begin{tabular}[c]{@{}l@{}}39.52dB\end{tabular}\\ 
\hline
SSIM & \begin{tabular}[c]{@{}l@{}}0.76\end{tabular} 
& \begin{tabular}[c]{@{}l@{}}0.95\end{tabular}\\
\bottomrule
\end{tabular}
\end{table}

Experimental results show that our proposed reconstruction method also has great performance on our acquired industrial CT image, which means our network has great generalization. Therefore, the image reconstruction quality of proposed system can meet the demand of industrial non-destructive testing.

\subsection{Time consumption analysis}\label{sec4}
\begin{figure}[htb]
    \centering
    \includegraphics[width=12cm]{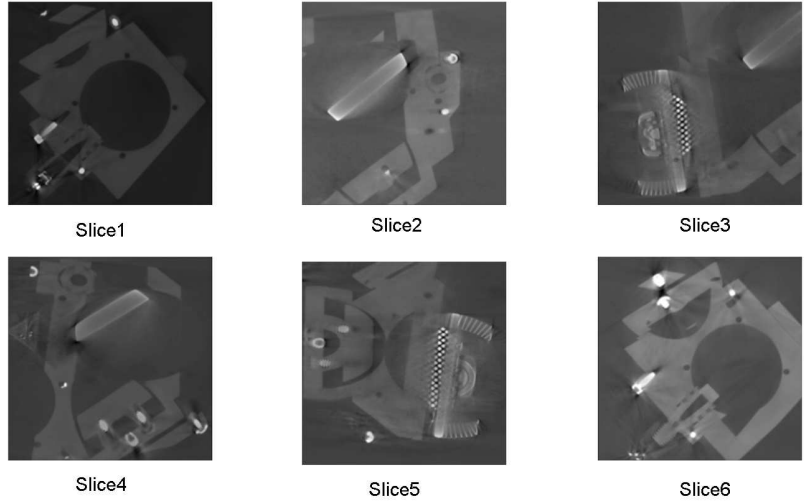}
    \caption{Slices for time consumption test}
    \label{Fig13}
\end{figure}
As is shown in figure13, we selected different slices of the spectral interferometer for reconstruction to calculate the reconstruction time consumption after these slices pass through the detector scan plane. Time consumption is mainly composed of two parts: sparse-view FBP process in cluster workstation and the post-processing in graphics workstation. We calculated the reconstruction time consumption of the two steps on the tested slices, the result was shown in table5.

\begin{table}[ht]
\centering
\caption{Time consumption of different slices}\label{tab5}%
\begin{tabular}{@{}cccccccc@{}}
\toprule
 & \begin{tabular}[c]{@{}l@{}}Slice1\end{tabular}
 & \begin{tabular}[c]{@{}l@{}}Slice2\end{tabular}
 & \begin{tabular}[c]{@{}l@{}}Slice3\end{tabular}
 & \begin{tabular}[c]{@{}l@{}}Slice4\end{tabular}
 & \begin{tabular}[c]{@{}l@{}}Slice5\end{tabular} 
 & \begin{tabular}[c]{@{}l@{}}Slice6\end{tabular} 
 & \begin{tabular}[c]{@{}l@{}}average\end{tabular} 
 \\
\hline
FBP/s & \begin{tabular}[c]{@{}l@{}}0.004\end{tabular} 
& \begin{tabular}[c]{@{}l@{}}0.004\end{tabular}
& \begin{tabular}[c]{@{}l@{}}0.004\end{tabular}
& \begin{tabular}[c]{@{}l@{}}0.004\end{tabular}
& \begin{tabular}[c]{@{}l@{}}0.005\end{tabular}
& \begin{tabular}[c]{@{}l@{}}0.004\end{tabular}
& \begin{tabular}[c]{@{}l@{}}0.004\end{tabular}
\\ 
\hline
Post-processing/s & \begin{tabular}[c]{@{}l@{}}0.125\end{tabular} 
& \begin{tabular}[c]{@{}l@{}}0.123\end{tabular}
& \begin{tabular}[c]{@{}l@{}}0.123\end{tabular}
& \begin{tabular}[c]{@{}l@{}}0.124\end{tabular}
& \begin{tabular}[c]{@{}l@{}}0.123\end{tabular}
& \begin{tabular}[c]{@{}l@{}}0.123\end{tabular}
& \begin{tabular}[c]{@{}l@{}}0.124\end{tabular}
\\ 
\hline
Total/s & \begin{tabular}[c]{@{}l@{}}0.129\end{tabular} 
& \begin{tabular}[c]{@{}l@{}}0.127\end{tabular}
& \begin{tabular}[c]{@{}l@{}}0.127\end{tabular}
& \begin{tabular}[c]{@{}l@{}}0.128\end{tabular}
& \begin{tabular}[c]{@{}l@{}}0.128\end{tabular}
& \begin{tabular}[c]{@{}l@{}}0.127\end{tabular}
& \begin{tabular}[c]{@{}l@{}}0.128\end{tabular}
\\
\bottomrule
\end{tabular}
\end{table}

In our proposed reconstruction solution, trained network is stored in the graphics workstation. Sparse-view projections acquired by the detectors were sent to the GPU in the cluster workstation for FBP reconstruction at first, then the reconstructed images were sent to the network as inputs, the final reconstructed images were acquired as the outputs of the network. 
We can see from table 5 that the time consumption is mainly determined by post-processing time, in our Nvidia RTX3060 GPU, time consumption of the whole procedure is around 0.128 seconds. We have reason to believe that time consumption can be decreased using more advanced GPU. By changing the interval between inspected objects on the convey belt, our system can meet the demands of real-time imaging. The gap between inspected objects should satisfy the following formula:

\begin{equation}
    g>S\cdot {{R}_{T}}
\end{equation}

In Eq. (17), g refers to gap between inspected objects, S refers to line speed of the convey belt, ${{R}_{T}}$ refers to the reconstruction time we estimated.
The readout time of the detector is crucial for the whole imaging process, if the readout time is too long, projections in the detectors cannot be guaranteed to be came from the same slice, which makes an enormous alignment error.

\section{Discussion and Conclusion}\label{sec5}
In our simulation, the interested slices of the inspected object were successfully constructed in a short time by the CT structure we proposed. The inspection part is highly integrated to the whole production line so that the quality inspection has been simplified and the production cycle has been greatly shortened. Meanwhile, sufficient quantity of X-ray detectors and the deep-learning reconstruction method make sure that the quality of reconstructed slices satisfies the needs of defect detection.

By adjusting the key parameters of the system, the real-time imaging frame rate and image quality are variable. Slices of comprehensive structure and dangerous section can be scanned with higher frame rate and imaging quality, while some irrelevant slices can be scanned with lower frame rate and imaging quality. Computing power is effectively saved in this arrangement. Some other solutions can also apply in our system, such as inspecting the object in the region of interest so that the reconstruction time can be decreased \cite{bib41}.

This is the first time that training the network using medical dataset and applying it to industrial CT reconstruction. The medical CT data can be viewed as source domain and the acquired industrial CT data can be viewed as target domain. Transfer learning can greatly reduce the training time and deployment cycle by applying the pre-trained model from the source domain to the target domain.

In the further study, we will focus on reducing the needed projection number for reconstruction by optimizing the deep-learning reconstruction network and applying the knowledge of transfer learning \cite{bib42}. In this way, number of X-ray source and detector can be further reduced, the system structure complexity and the cost will be reduced as well.

With the guarantee of imaging quality and imaging speed, our proposed system provides an ideal solution for the sorting, grading, quality inspection, material analysis, and optimization of complex manufacturing processes in recycling, food processing, mining, and other process industries. It can also be applied in the field of security as well.

Our proposed system can significantly increase the speed and accuracy of inspection in production process and it is convenient to be integrated into production lines. Evaluation metrics and inspection images are updated in real time, the quality of each product can be verified, not just spot checked. It provides an ideal solution for quality inspection visualization in the age of Industry 4.0.

\section{Acknowledgement}\label{sec6}
We would like to thank for the support of National Natural Science Foundation of China (61571381) and Double-Hundred Project for the Introduction of Xiamen Elite (Special Talent 2019.1). We also want to thank the Research Center of Aircraft Health Management Technology of Xiamen University for some constructive comments.

\bibliographystyle{unsrt}  
\bibliography{references}  

\begin{thebibliography}{10}

\bibitem{bib1}
Wilhelm~Conrad. Roentgen.
\newblock On a new kind of rays.
\newblock {\em CA: a cancer journal for clinicians}, 22(3):153--157, 1972.

\bibitem{bib2}
Hounsfield~G N.
\newblock Computerized transverse axial scanning (tomography): Part 1.
  description of system[j].
\newblock {\em The British journal of radiology}, 46(552):1016--1022, 1973.

\bibitem{bib3}
Min Yang and et~al.
\newblock Extra projection data identification method for
  fast-continuous-rotation industrial cone-beam ct.
\newblock {\em Journal of X-ray Science and Technology}, 21(4):467--479, 2013.

\bibitem{bib4}
Guorui Yan and et~al.
\newblock Fast cone-beam ct image reconstruction using gpu hardware.
\newblock {\em Journal of X-ray Science and Technology}, 16(4):225--234, 2008.

\bibitem{bib5}
Federico Giudiceandrea, Ursella Enrico, and Enrico Vicario.
\newblock A high speed ct scanner for the sawmill industry., 2011.
\newblock Proceedings of the 17th international non destructive testing and
  evaluation of wood symposium. Sopron, Hungary: University of West
  Hungary,2011.

\bibitem{bib6}
Ursella Enrico, Federico Giudiceandrea, and Marco Boschetti.
\newblock A fast and continuous ct scanner for the optimization of logs in a
  sawmill., 2018.
\newblock Paper presented at the 8th Conference on Industrial Computed
  Tomography (iCT 2018) at Wels, Austria. Vol. 2. 2018.

\bibitem{bib7}
Robb~Richard A. and et~al.
\newblock High-speed three-dimensional x-ray computed tomography: The dynamic
  spatial reconstructor., 1983.
\newblock Proceedings of the IEEE 71.3 (1983): 308-319.

\bibitem{bib8}
Weiwen Wu and et~al.
\newblock Ai-enabled ultra-low-dose ct reconstruction., 2021.

\bibitem{bib9}
Sidky~Emil Y. and et~al.
\newblock Do cnns solve the ct inverse problem?
\newblock {\em IEEE Transactions on Biomedical Engineering}, 68(8):1799--1810,
  2020.

\bibitem{bib10}
Han Yoseob and Jong~Chul Ye.
\newblock Framing u-net via deep convolutional framelets: Application to
  sparse-view ct.
\newblock {\em IEEE transactions on medical imaging}, 37(6):1418--1429, 2018.

\bibitem{bib11}
Zhang Hanming and et~al.
\newblock Image prediction for limited-angle tomography via deep learning with
  convolutional neural network., 2016.

\bibitem{bib12}
Van~Tiggelen R.
\newblock In search for the third dimension: from radiostereoscopy to
  three-dimensional imaging[j].
\newblock {\em JBR-BTR}, 85(5):266--270, 2002.

\bibitem{bib13}
Thomas~Adrian MK and Banerjee~Arpan K.
\newblock {\em The history of radiology}.
\newblock OUP Oxford, New York, 2013.

\bibitem{bib14}
Webb Steve.
\newblock {\em From the watching of shadows: the origins of radiological
  tomography}.
\newblock CRC Press, Boca Raton, 1990.

\bibitem{bib15}
Kevles Bettyann.
\newblock {\em Naked to the bone: Medical imaging in the twentieth century}.
\newblock Rutgers University Press, Chicago, 1997.

\bibitem{bib16}
Moore TD, Vanderstraeten D, and Forssell P.
\newblock Determination of bga structural defects and solder joint defects by
  3d x-ray laminography., 2001.
\newblock Proceedings of the 2001 8th International Symposium on the Physical
  and Failure Analysis of Integrated Circuits. IPFA 2001 (Cat. No. 01TH8548):
  146--150.

\bibitem{bib17}
Adam Thompson and Richard. Leach.
\newblock Introduction to industrial x-ray computed tomography.
\newblock In {\em Industrial X-ray Computed Tomography}, pages 1--23. Springer,
  New York, 2018.

\bibitem{bib18}
WB~Gilboy and J.~Foster.
\newblock Industrial applications of computerized tomography with x-and
  gamma-radiation.
\newblock In {\em Research techniques in nondestructive testing vol. 6}, pages
  255--287. Academic P, London (UK), 1982.

\bibitem{bib19}
Peter Reimers and J{\"u}rgen. Goebbels.
\newblock New possibility of nondestructive evaluation by x-ray computed
  tomography.
\newblock {\em Materials evaluation}, 41:732--737, 1983.

\bibitem{bib20}
JW~Kress and LA. Feldkamp.
\newblock X-ray tomography applied to nde of ceramics., 1983.
\newblock ASME 1983 International Gas Turbine Conference and Exhibit.

\bibitem{bib21}
Reinhold Oster.
\newblock Computed tomography as a nondestructive test method for fiber main
  rotor blades in development, series and maintenance, 1997.
\newblock European Rotorcraft Forum, 23 rd, Dresden, Germany.

\bibitem{bib22}
C.~Liu, R.R. Wang, I.~Ho, and et~al.
\newblock Toward online layer-wise surface morphology measurement in additive
  manufacturing using a deep learning-based approach.
\newblock {\em Journal of Intelligent Manufacturing}, pages 1--17, 2022.

\bibitem{bib23}
Mohamed Elhefnawy, Ahmed Ragab, and Mohamed-Salah. Ouali.
\newblock Fault classification in the process industry using polygon generation
  and deep learning.
\newblock {\em Journal of Intelligent Manufacturing}, 33(5):1531--1544, 2022.

\bibitem{bib24}
Zhuxi Ma, Yibo Li, Minghui Huang, Qianbin Huang, Jie Cheng, and Si. Tang.
\newblock Automated real-time detection of surface defects in manufacturing
  processes of aluminum alloy strip using a lightweight network architecture.
\newblock {\em Journal of Intelligent Manufacturing}, pages 1--17, 2022.

\bibitem{bib25}
Monica~L Nogueira, Noel~P Greis, Rachit Shah, Matthew~A Davies, and Nicholas~E.
  Sizemore.
\newblock Machine learning classification of surface fracture in
  ultra-precision diamond turning using csi intensity map images.
\newblock {\em Journal of Manufacturing Systems}, 2022.

\bibitem{bib26}
Chuqiao Xu, Junliang Wang, Jing Tao, Jie Zhang, and Pai. Zheng.
\newblock A knowledge augmented deep learning method for vision-based yarn
  contour detection.
\newblock {\em Journal of Manufacturing Systems}, 63:317--328, 2022.

\bibitem{bib27}
A.~Cramer, J.~Hecla, D.~Wu, and et~al.
\newblock Stationary computed tomography for space and other
  resource-constrained environments.
\newblock {\em Scientific reports}, 8(1):1--10, 2018.

\bibitem{bib28}
Tao Zhang, Yuxiang Xing, Li~Zhang, Xin Jin, Hewei Gao, and Zhiqiang. Chen.
\newblock Stationary computed tomography with source and detector in linear
  symmetric geometry: Direct filtered backprojection reconstruction.
\newblock {\em Medical physics}, 47(5):2222--2236, 2020.

\bibitem{bib29}
Hongguang Cao, LI~Yunxiang, Tong Chang, Zhili Cui, and Hailiang Zheng.
\newblock Stationary real time ct imaging system and method thereof, August~18
  2020.
\newblock US Patent 10,743,826.

\bibitem{bib30}
Derrek Spronk, Yueting Luo, Christina~R Inscoe, Yueh~Z Lee, Jianping Lu, and
  Otto. Zhou.
\newblock Evaluation of carbon nanotube x-ray source array for stationary head
  computed tomography.
\newblock {\em Medical physics}, 48(3):1089--1099, 2021.

\bibitem{bib31}
Xin Qian, Andrew Tucker, Emily Gidcumb, Jing Shan, Guang Yang, Xiomara
  Calderon-Colon, Shabana Sultana, Jianping Lu, Otto Zhou, Derrek Spronk, and
  others.
\newblock High resolution stationary digital breast tomosynthesis using
  distributed carbon nanotube x-ray source array.
\newblock {\em Medical physics}, 39(4):2090--2099, 2012.

\bibitem{bib32}
Rafael~C Gonzalez.
\newblock {\em Digital image processing}.
\newblock Pearson education india, Noida, 2009.

\bibitem{bib33}
Wim Van~Aarle, Willem~Jan Palenstijn, Jeroen Cant, Eline Janssens, Folkert
  Bleichrodt, Andrei Dabravolski, Jan De~Beenhouwer, K~Joost Batenburg, and
  Jan. Sijbers.
\newblock Fast and flexible x-ray tomography using the astra toolbox.
\newblock {\em Optics express}, 24(22):25129--25147, 2016.

\bibitem{bib34}
Hu~Chen, Yi~Zhang, Mannudeep~K Kalra, Feng Lin, Yang Chen, Peixi Liao, Jiliu
  Zhou, and Ge. Wang.
\newblock Low-dose ct with a residual encoder-decoder convolutional neural
  network.
\newblock {\em IEEE transactions on medical imaging}, 36(12):2524--2535, 2017.

\bibitem{bib35}
Kyong~Hwan Jin, Michael~T McCann, Emmanuel Froustey, and Michael. Unser.
\newblock Deep convolutional neural network for inverse problems in imaging.
\newblock {\em IEEE Transactions on Image Processing}, 26(9):4509--4522, 2017.

\bibitem{bib36}
Qingsong Yang, Pingkun Yan, Yanbo Zhang, Hengyong Yu, Yongyi Shi, Xuanqin Mou,
  Mannudeep~K Kalra, Yi~Zhang, Ling Sun, and Ge. Wang.
\newblock Low-dose ct image denoising using a generative adversarial network
  with wasserstein distance and perceptual loss.
\newblock {\em IEEE transactions on medical imaging}, 37(6):1348--1357, 2018.

\bibitem{bib37}
Olaf Ronneberger, Philipp Fischer, and Thomas. Brox.
\newblock U-net: Convolutional networks for biomedical image segmentation.,
  2015.
\newblock International Conference on Medical image computing and
  computer-assisted intervention.

\bibitem{bib38}
Jianjie Lu and Kai-yu. Tong.
\newblock Visualized insights into the optimization landscape of fully
  convolutional networks., 2019.

\bibitem{bib39}
Kaiming He, Xiangyu Zhang, Shaoqing Ren, and Jian. Sun.
\newblock Deep residual learning for image recognition., June 2016.
\newblock Proceedings of the IEEE conference on computer vision and pattern
  recognition.

\bibitem{bib40}
J.~Leuschner, M.~Schmidt, D.O. Baguer, and et~al.
\newblock Lodopab-ct, a benchmark dataset for low-dose computed tomography
  reconstruction.
\newblock {\em Scientific Data}, 8(1):1--12, 2021.

\bibitem{bib41}
P~Reimers, A~Kettschau, and J.~Goebbels.
\newblock Region-of-interest (roi) mode in industrial x-ray computed
  tomography.
\newblock {\em NDT international}, 23(5):255--261, 1990.

\bibitem{bib42}
Sinno~Jialin Pan and Qiang. Yang.
\newblock A survey on transfer learning.
\newblock {\em IEEE Transactions on knowledge and data engineering},
  22(10):1345--1359, 2009.

\end{thebibliography}






\end{document}